\documentclass[aps,prb,twocolumn,superscriptaddress,amsmath]{revtex4-1}

\usepackage{graphicx,subfig,bm,wasysym,color}
\usepackage{amsmath,amsfonts,amsthm,amssymb,amsbsy,bbold,mathrsfs}
\definecolor{orange}{rgb}{1,0.5,0}
\newcommand{\beq}{\begin{equation}}
\newcommand{\eeq}{\end{equation}}
\newcommand{\bea}{\begin{eqnarray}}
\newcommand{\eea}{\end{eqnarray}}
\newcommand{\beqa}{\begin{eqnarray}}
\newcommand{\eeqa}{\end{eqnarray}}

\newcommand{\HKH}{$H_\text{KH}$ }
\newcommand{\eqh}{Eq.~\ref{eq:KH} }

\begin{document}
\title{Kitaev-Heisenberg models for iridates on the triangular, hyperkagome, kagome, fcc, and pyrochlore lattices} 
\author{Itamar Kimchi}
\affiliation{Department of Physics, University of California, Berkeley, CA 94720, USA}
\author{Ashvin Vishwanath}
\affiliation{Department of Physics, University of California, Berkeley, CA 94720, USA}
\affiliation{Materials Science Division, Lawrence Berkeley National Laboratories, Berkeley, CA 94720, USA}
\date{\today}
\begin{abstract}
The Kitaev-Heisenberg (KH) model has been proposed to capture magnetic interactions in iridate Mott insulators on the honeycomb lattice.
We show that analogous interactions arise in many other geometries built from edge-sharing IrO$_6$ octahedra,
including the pyrochlore and hyperkagome lattices relevant to Ir$_2$O$_4$ and Na$_4$Ir$_3$O$_8$ respectively. The Kitaev spin liquid exact solution does not generalize to these lattices.
However, a different exactly soluble point of the honeycomb lattice KH model, obtained by a four-sublattice transformation to a ferromagnet, generalizes to all of these lattices and even to certain additional further neighbor Heisenberg couplings. A Klein four-group $\cong \mathbb{Z}_2 \times \mathbb{Z}_2$ structure is associated with this mapping  (hence \textit{Klein duality}). A finite lattice admits the duality if a simple geometrical condition is met. This duality predicts fluctuation free ordered states on these different 2D and 3D lattices, which are analogs of the honeycomb lattice KH \textit{stripy} order.
 This result is used in conjunction with a semiclassical  Luttinger-Tisza approximation to obtain phase diagrams for KH models on the different lattices. We also discuss a Majorana fermion based mean field theory at the Kitaev point, which is exact on the honeycomb lattice,  for the KH models on the different lattices. We attribute the rich behavior of these models to the interplay of geometric frustration and frustration induced by spin-orbit coupling.
\end{abstract}
\maketitle
\section{Introduction}
Long viewed as a perturbative correction, relativistic spin-orbit coupling has in recent years been increasingly asserting its role within condensed matter physics.  It took center stage with topological insulators, time reversal invariant states of electrons with no strong interactions that use spin-orbit coupling to generate nontrivial topology in the band structure\cite{Zhang2011,Moore2011,Kane2010}.  Electron correlations may amplify\cite{Balents2010b,Savrasov2011} the effects of spin-orbit coupling (SOC), enriching the taxonomy of possible phases. Thus Mott insulating states of heavy magnetic ions could realize novel Hamiltonians, which may be hitherto unexplored or not thought to describe real materials.

One such S=1/2 Hamiltonian has been proposed by Jackelli and Khaliullin\cite{Khaliullin2009,Khaliullin2010} to occur in the honeycomb iridates Na$_2$IrO$_3$ and Li$_2$IrO$_3$.
It includes the Kitaev exchange, a nearest neighbor ising coupling of spin component $\gamma\in\{x,y,z\}$ set by the spatial orientation of the bond\cite{Khaliullin2001,Kitaev2006}.  The pure Kitaev honeycomb Hamiltonian is exactly solvable with a quantum spin liquid (QSL) ground state of a gapless Majorana coupled to $\mathbb{Z}_2$ fluxes\cite{Kitaev2006}. The proposed magnetic model for these iridates includes the Kitaev as well as SU(2) symmetric Heisenberg coupling, yielding the Kitaev-Heisenberg $S=1/2$ Hamiltonian\cite{Khaliullin2010}. It may be written as
\beq
H_{\text{KH}} =
\sum_{\left< i j\right>}\eta \left[\left( 1-\left|\alpha\right| \right)\vec{S}_i\cdot\vec{S}_j -2\alpha S^{\gamma_{ij}}_i S^{\gamma_{ij}}_j \right]
\label{eq:KH}
\eeq
with
 $\eta=\pm 1$ and $-1 \leq \alpha \leq 1$. Here $\eta$ sets the sign of Heisenberg exchange, and negative $\alpha$ gives the same sign for both exchanges. Pairs of endpoints of the two $\alpha$-segments for $\eta=+1,-1$ are identified as a single point by the product $\eta \alpha=+1$ (FM Kitaev) and similarly $\eta \alpha=-1$ (AF Kitaev), forming an $(\eta,\alpha)$ parameter ring\cite{Khaliullin2012}. We will primarily focus on this idealized Hamiltonian but also consider some extensions such as farther neighbor couplings.

The phase diagram of \HKH on the honeycomb lattice is known from a combination of exact diagonalization\cite{Khaliullin2010,Khaliullin2012}, other numerical methods\cite{Trebst2011,Trebst2011b} and the presence of exactly soluble points. In addition to the exact solution using majorana fermions of the Kitaev Hamiltonians $\alpha=\pm 1$, and the obvious SU(2)-symmetric ferromagnet ($\eta=-1,\alpha=0$), a four sublattice site-dependent spin rotation\cite{Khaliullin2010} transforms \HKH at $\eta=+1,\alpha=1/2$ into a ferromagnet in the rotated basis.  The original spins are then ``stripy'' ordered. Neel order from the Heisenberg antiferromagnet is unfrustrated on the bipartite honeycomb, and
was recently shown\cite{Khaliullin2012} to map under this transformation to a physical parameter regime hosting the spin pattern known as ``zigzag''.

This zigzag phase was determined in recent experiments\cite{Taylor2012,Hill2011,Cao2012} to be the low temperature ordering pattern of Na$_2$IrO$_3$. The zigzag order was earlier theoretically found to be most stabilized by combining Kitaev interactions and the further neighbor\cite{Lauchli2011,Thomale2011} exchanges $J_2$, $J_3$ which naturally arise across a honeycomb hexagon\cite{Khomskii2012,Valenti2013}, and which together fit the available experimental data in comparisons to exact diagonalization\cite{You2011}.
Indeed, within a classical approximation to the phase diagram (Luttinger-Tisza described below), the zigzag phase within the pure Kitaev-Heisenberg model lies nearly at the boundary of the large zigzag-ordered region stabilized by $J_2, J_3$ exchanges\cite{unpublished}. Interestingly, the zigzag phase in its $J_1-J_2-J_3$ limit and in its Eq.~\ref{eq:KH} limit may offer experimentally relevant distinguishing characteristics\cite{Khaliullin2012,Horsch2013}.

So far only the honeycomb iridates Na$_2$IrO$_3$ and Li$_2$IrO$_3$ have been studied in the context of the Kitaev-Heisenberg model. Despite initial worries that trigonal distortion would invalidate the derivation of the Kitaev exchange discussed below, recent resonant inelastic x-ray scattering results\cite{Kim2012} support the validity of the strong spin-orbit coupling approach. For the sodium iridate Na$_2$IrO$_3$, attempts to extract the magnetic Hamiltonian from fits to experiments including susceptibility and spin wave spectra\cite{Gegenwart2010,Gegenwart2012,Taylor2012,You2011,Hill2011,Cao2012,Khaliullin2012} and to electronic properties\cite{Damascelli2012,Horsch2013} have so far proved unable to distinguish between substantial Kitaev exchange and a complete lack of it. Few experimental results on magnetic behavior in the lithium iridate Li$_2$IrO$_3$ are currently available, though the relatively small magnitude of the Curie Weiss scale extracted from susceptibility suggests the Kitaev exchange may be strong\cite{Gegenwart2012,You2011}.

Beyond the possible Kitaev-Heisenberg physics in the layered honeycomb iridates, other
iridates have also attracted much attention.
Layered compounds include the Mott insulator
Sr$_2$IrO$_4$\cite{Arima2009,Rotenberg2008} and its bilayer variant
Sr$_3$Ir$_2$O$_7$\cite{Marshall2002,Noh2008}, both with ordered SOC magnetic moments.
 Notable examples with a fully three dimensional structure include the 2-2-7 pyrochlore iridates, where changing the $A$ site rare earth metal yields radically varying properties\cite{Maeno2001,Savrasov2011,Onoda2007,Balents2010b}; the sodium iridate Na$_4$Ir$_3$O$_8$ spin liquid candidate, with Ir on the pyrochlore-descendent hyperkagome lattice\cite{Takagi2007}; and a recently epitaxially-stabilized iridium spinel Ir$_2$O$_4$ with empty cation sites\cite{Takagi2010} leaving Ir on a pyrochlore lattice. Despite the variety of elemental composition and geometrical structure in this list, there is a simple but fundamental distinction separating the latter two compounds from the others listed.

\begin{figure*}[htp]
\includegraphics[scale=0.59]{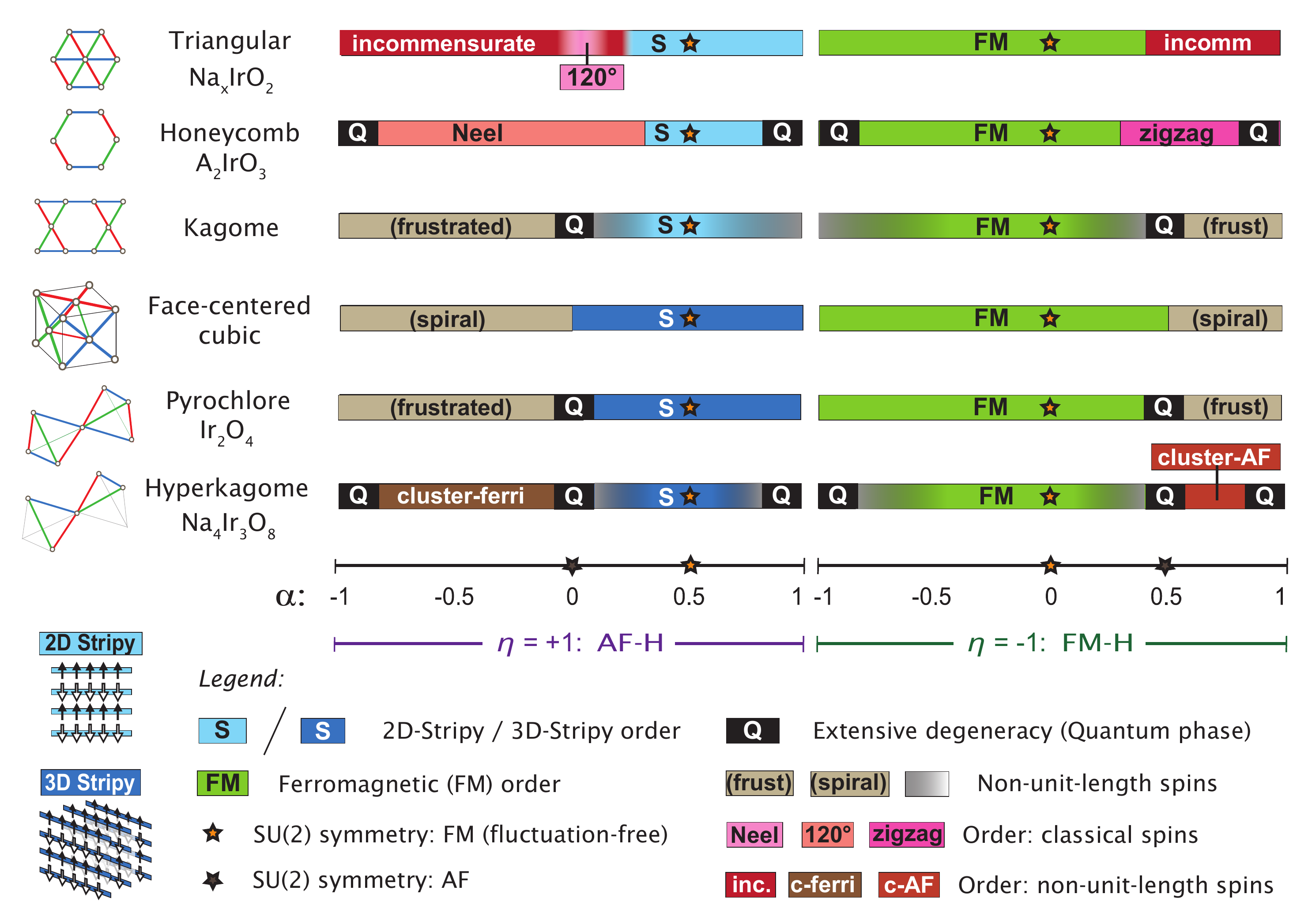}
\caption{{\bf Phase diagrams using the Klein duality as well as the Luttinger-Tisza approximation (LTA).}
Phase diagrams are shown for the Kitaev-Heisenberg Hamiltonian,
$H_{KH} = \eta \left( 1-\left|\alpha\right| \right)\vec{S}_i\cdot\vec{S}_j -2\eta\alpha S^{\gamma_{ij}}_i S^{\gamma_{ij}}_j  $ with $\eta=\pm 1$, on the various lattices. Note that the parameter space is a ring: for the $\alpha$ parameter segments shown here the endpoints should be identified, i.e. writing the parameter as $(\eta,\alpha)$, identify the points (-1,-1)$\cong$(+1,+1) and also identify the points (-1,+1)$\cong$(+1,-1). Rich phase diagrams are found. 
2D and 3D stripy magnetic phases (blue), found on all lattices, are exact and fluctuation free by the Klein duality at the FM SU(2) point (yellow star) $\eta$=$+1,\alpha$=$1/2$.
On the kagome and hyperkagome lattices, outside the FM SU(2) points, the FM and stripy orders are given
non-unit-length spins by the LTA (gray shading), suggesting frustration.
Extensive degeneracy of LTA ordering wavevectors hints at a non-spin-ordered ``quantum'' phase, labelled "$Q$": where the Hamiltonians hosting $Q$ phases have been solved, exactly\cite{Kitaev2006} for the honeycomb at $\alpha$=$\pm1$ and numerically\cite{White2011,Balents2012} for the kagome at $\eta$=$+1,\alpha$=$0$, they have turned out to host non-magnetic phases. The hyperkagome hosts $Q$ points at the Heisenberg antiferromagnet and its Klein dual, as well as at the pure Kitaev Hamiltonians, because any single Kitaev bond type fragments the hyperkagome into disjoint clusters (Fig.~\ref{fig:hyperkagome-unit-cell}). The  $120^\circ$ triangular lattice and Neel and zigzag honeycomb orders are found with normalized spins. Apparent ordering with LTA non-normalized spins is found at incommensurate wavevectors on the triangular lattice and in the cluster-ferrimagnet and cluster-antiferromagnet (AF) regimes on the hyperkagome. Kagome, fcc and pyrochlore lattices also host frustrated regimes with no definitive ordering within the LTA, as described in the text (gray).
 }
\label{fig:O3-diagrams}
\end{figure*}

In this manuscript we show that the iridates Na$_4$Ir$_3$O$_8$ and Ir$_2$O$_4$, as well as possible compounds in certain other geometries, may be described by generalizations of the Hamiltonian Eq.~\ref{eq:KH} to the relevant lattices (hyperkagome for Na$_4$Ir$_3$O$_8$ and pyrochlore for Ir$_2$O$_4$). The key quantum chemistry ingredients which can generate the interactions \HKH have been already pointed out by Jackeli and Khaliullin\cite{Khaliullin2009} but the extension to three dimensional lattices, as well as to these compounds, has not been previously exposed. We begin by recalling the derivation of \HKH and systematically extending it to other geometries in two and three dimensions; it applies when oxygen octahedra are edge-sharing, yielding lattices that are in a certain sense subsets of the fcc. We then proceed to investigate the phase diagram of \HKH on these lattices, using primarily analytical approaches.
We generalize the honeycomb four-sublattice transformation into a duality on graphs and lattices with Kitaev labeled bonds in any dimension, and even with certain further neighbor pure Heisenberg couplings; we shall refer to it as the \textit{Klein duality} since, as we shall show, it is structured by the Klein four-group $\cong \mathbb{Z}_2 \times \mathbb{Z}_2$. We give a simple algorithm determining which graphs admit the duality, based on this Klein group structure. The Klein duality gives stripy phases as FM-duals.
Diagonalizing the classical version of \HKH with spins of unconstrained length (i.e. the  Luttinger-Tisza approximation), we identify unconventional ordering patterns and also find hints of quantum magnetically disordered phases, most interestingly on the hyperkagome. The Luttinger-Tisza phase diagrams are shown in Fig.~\ref{fig:O3-diagrams}.
To directly capture Majorana fermion quantum spin liquids analogous to the Kitaev honeycomb QSL, we decompose spins into Majorana combinations of Schwinger fermions, a mean field treatment which is exact for the Kitaev honeycomb model,
 finding on all other lattices fermionic QSLs which break time reversal and carry gapless excitations, but which are not exact solutions.
 For the honeycomb and hyperkagome pure Kitaev Hamiltonians, the lattice fragments under a bond type $\gamma$ into disjointed localized clusters, giving flat bands in the Majorana mean field as well as in the Luttinger-Tisza approximation, which hints at a possible analogy between the honeycomb Kitaev QSL and the Kitaev Hamiltonian ground state on the hyperkagome.

We focus on two candidate materials, while also considering other related compounds.
The recently epitaxially fabricated Ir$_2$O$_4$ is a spinel without the $A$ cation, leaving the iridium ions on a pyrochlore lattice with oxygens positioned appropriately, as described below; Ir$_2$O$_4$ was found to be a narrow gap insulator\cite{Takagi2010}.  The spin liquid candidate Na$_4$Ir$_3$O$_8$ is an iridate with $S=1/2$ moments on the three dimensional hyperkagome lattice, which exhibits no magnetic order down to at least 2K\cite{Takagi2007}.
In addition to these two iridates, this study may also capture compounds in which iridium is replaced by a transition metal ion with strong spin orbit coupling, intermediate correlations and valency appropriate for a magnetic effective spin one half model (see below). Recently, the osmium oxides CaOs$_2$O$_4$ and SrOs$_2$O$_4$ were computationally predicted\cite{Savrasov2012} to be stabilized in the spinel structure relevant to Kitaev-Heisenberg physics;
if they indeed exist in this geometry,
\HKH should form at least part of their magnetic Hamiltonian.
Kagome and triangular lattice iridates may be seen as appropriate layers within epitaxially stabilized Ir$_2$O$_4$, coupled together in a nontrivial manner.  Triangular lattice iridates could potentially also be stabilized as analogues of the cobaltates\cite{Blanchard2008} Na$_x$CoO$_2$, where Co is on a triangular lattice; the preferred valency would exist uniformly only in the limit $x\rightarrow 0$, but small $x$ should offer interesting perturbations as well as likely separate layers of triangular lattices.  However, no triangular lattice iridate with the relevant edge sharing octahedra structure is currently available; a compound of the type Na$_x$IrO$_2$ may or may not turn out to be stabilized.

In addition to the honeycomb Kitaev-Heisenberg model, other previous work has investigated Hamiltonians related to \HKH on other lattices.
Chen and Balents\cite{Balents2008} studied spin Hamiltonians on the hyperkagome lattice for Na$_4$Ir$_3$O$_8$, in the strong and weak SOC limits (relative to octahedral distortions). Within the strong SOC case they considered the single superexchange pathway via oxygen ions generating the single point \HKH  at $\alpha=1/2$, for which they found that classical configurations of stripy patterns were completely unfrustrated
\footnote{We now know that even the \textit{quantum} Hamiltonian is exactly soluble with its ground state free of fluctuations, via the Klein duality.}.
Superexchange via oxygen ions generating anisotropic spin interactions for the SOC Kramers doublet in Na$_4$Ir$_3$O$_8$ was also considered by Micklitz and Norman\cite{Micklitz2010,Norman2010}
in electronic structure computations and associated microscopic tight binding parametrization. Recently Reuther, Thomale and Rachel\cite{Rachel2012} studied a family of related Hamiltonians including on triangular lattices formed by second neighbors of the honeycomb, with an associated hidden ferromagnet.  
Very recently the classical \HKH Hamiltonian was studied on the triangular lattice by
Rousochatzakis, Rossler, Brink and Daghofer\cite{Daghofer2012}. Their study included a classical Monte Carlo computation suggesting the intriguing possibility that Kitaev exchange can stabilize an incommensurate vortex lattice of the $\mathbb{Z}_2$ topological defects of the Heisenberg antiferromagnet $120^\circ$ order.

\section{Kitaev couplings in lattices beyond the honeycomb}
\begin{figure}[tb]
\includegraphics[scale=0.4]{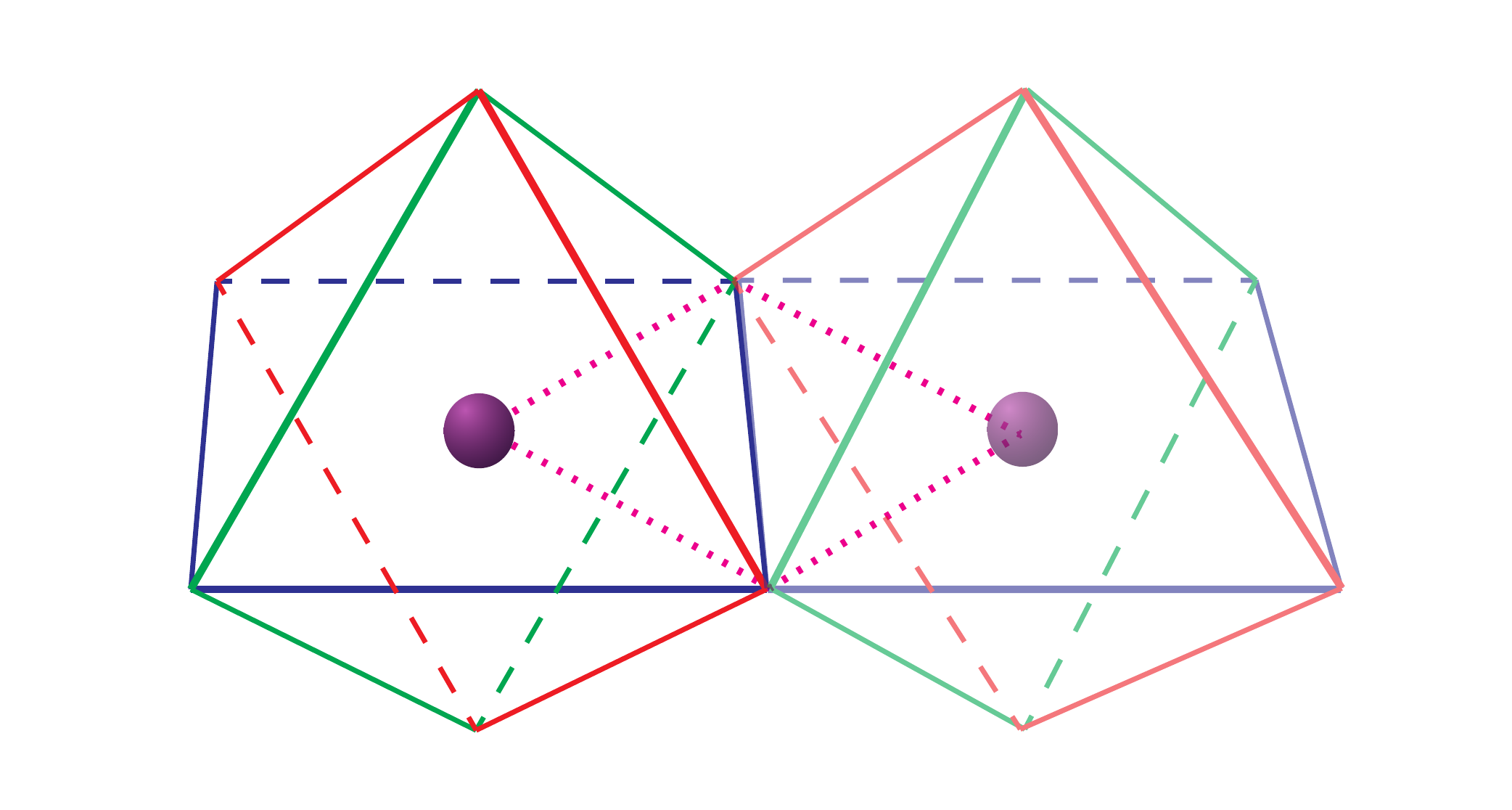}
\caption{{\bf Edge-sharing IrO$_6$ Octahedra generating Kitaev exchange.} Iridium ions (spheres) are each coordinated by six oxygen ions forming vertices of octahedra. Octahedra of neighboring Ir ions share edges. Dotted (purple) lines show the iridium-oxygen-iridium hopping paths, which form a square with $90^{\circ}$ angles. As described in the text, these superexchange paths generate an ising interaction between the iridium effective spins, which couple a spin component $x$, $y$ or $z$ depending on the orientation of the shared octahedra edge (shown in red, green and blue). Ir lattices hosting this Kitaev exchange must arise from a regular tiling of these edge-sharing octahedra.
\label{fig:octahedra} }
\end{figure}

Generating the Kitaev coupling requires a subtle recipe with ingredients from chemistry, geometry, and a hierarchy of energy scales, as we now recall\cite{Khaliullin2009,Khaliullin2010,Khaliullin2012}. Spin orbit coupling is key, together with (intermediate) correlations; let us focus on iridium.
The iridium ions should retain their 5d electrons in localized orbitals, and exist in the 4+ valence. 
Each iridium should be surrounded by six oxygen ions (or other electronegative ions with valence p-orbitals), which form the vertices of an octahedron cage, shown in Fig.~\ref{fig:octahedra}.
 The octahedral crystal field splits the 5d orbitals into an empty $e_g$ pair and a triplet of $t_{2g}$ orbitals with five electrons and one hole. Strong spin-orbit coupling further splits $t_{2g}$ down to a half-filled Kramer's doublet, the spin-1/2 degree of freedom defining the low energy manifold.
The final key ingredient is the geometrical structure: edge sharing octahedra with 90$^{\circ}$ Ir-O-Ir bond angles.

In perturbation theory from the Mott insulator limit, virtual hopping of holes from iridium $t_{2g}$ orbitals through intermediate oxygen $p$ orbitals generate the low energy spin Hamiltonian. There are multiple relevant exchange paths\cite{Khaliullin2010,Khaliullin2012}. When holes hop through intermediate oxygens and meet on an iridium d-orbital, the resulting coupling is a pure Kitaev term, and is proportional to $J_H / \left( (U_d-3J_H)(U_d-J_H) \right)\approx J_H/U_d^2$. The iridium Coulomb exchange $U_d$ and Hund's rule coupling $J_H$ together specify all of the multi-band interaction parameters, due to the symmetries of d-orbitals.
A second exchange path, with two holes meeting on an oxygen or cycling around the Ir-O square, contributes a combination of Kitaev and Heisenberg couplings equal to  \HKH  at $\alpha=1/2$, with a coefficient and sign $\eta$ depending on the oxygen p-orbitals charge-transfer gap and Coulomb repulsion. Direct iridium wavefunction overlap gives a pure Heisenberg coupling. Recently\cite{Khaliullin2012}, an additional pathway through the higher $e_g$ orbitals has been proposed to be relevant as well, contributing \HKH  at $\eta=-1$, $\alpha=1/2$. The interplay of these exchanges suggests $\alpha$ may not be computable microscopically.

Generalizing this derivation to geometries beyond the honeycomb requires preserving the edge sharing octahedra with 90$^{\circ}$ Ir-O-Ir bonds.
Many commonly studied iridates such as the layered perovskites and the ``2-2-7'' pyrochlores
have corner sharing octahedra and thus are not captured by this derivation. Fig.~\ref{fig:octahedra} shows two adjacent octahedra, with edges color coded by the spin component coupling they generate when the octahedra of neighboring Ir ions share that edge. It is evident that all twelve octahedra edges may be shared while still maintaining 90$^{\circ}$ bonds and three-fold symmetries (coupled space and spin rotations). Tiling octahedra which touch along edges builds a face-centered cubic (fcc) lattice of the octahedra centers.

\begin{figure}[tb]
\includegraphics[scale=0.59]{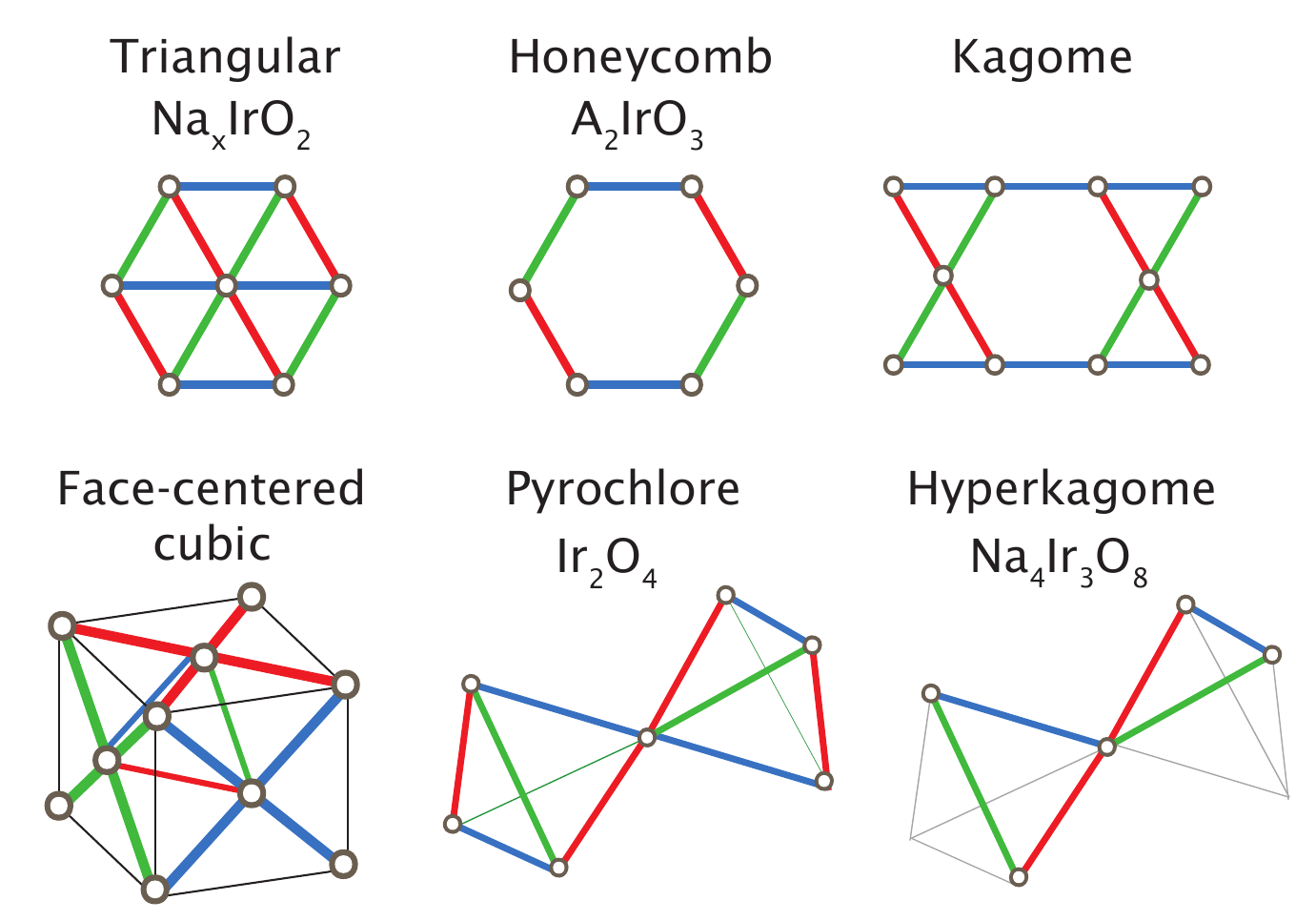}
\caption{{\bf Kitaev-Heisenberg lattices. \label{fig:KH-lattices}}
Iridium ions arranged in these lattices may generate the Kitaev spin exchange, coupling component $x,y,z$ on bonds colored red, green, blue respectively. A blue colored bond connecting two Ir sites implies that the respective IrO$^6$ octahedra share a blue edge as in Fig.~\ref{fig:octahedra}.
We also list examples of possible relevant iridium compounds which form these lattices.}
\end{figure}

We thus find that in two and three dimensions, all lattices whose graph of nearest neighbor bonds is a subset of the nearest neighbor bonds of the fcc, including the fcc itself, may host analogues of the Kitaev exchange. Possible geometries include the kagome and triangular lattices in two dimensions, and the face-centered cubic, pyrochlore (as realized in spinel-based compounds) and hyperkagome geometries in three dimensions. These are shown in Fig.~\ref{fig:KH-lattices}. These six are commonly studied lattices which are such subsets of the fcc, but an infinite number of lattices may be added to this list. 
The materials discussed above all have these edge-sharing octahedral structures and their magnetic Ir ions form one of these lattices. As for the honeycomb iridates, reduced crystal symmetry distorting Ir-O angles away from 90$^\circ$ will generate other magnetic exchanges.  Despite apparently strong $>10\%$ distortions of the bond angle in the sodium honeycomb iridate and a slew of experiments on this material, a Kitaev exchange comparable to or even stronger than the Heisenberg exchange is still consistent with current experimental results, suggesting a hopeful outlook for the other materials.

Note that the quantum chemistry considerations pictured in Fig.~\ref{fig:octahedra} tightly constrain the possible lattice realizations of Eq.~\ref{eq:KH}. Specifically, these constraints are \textit{tighter} than those imposed by naive symmetry considerations of SOC. For example, it is natural to define an implementation of SOC that couples spin component $S^z$ to bonds along $\hat{z}$, i.e. locks the Bloch sphere to real space. This would generate Eq.~\ref{eq:KH} on the simple cubic lattice with $S^zS^z$ coupling along $\hat{z}$ bonds, as well as on the square lattice with $\gamma=x,y$. But the exchange pathways of Ir $t_{2g}$ orbitals forbid this scenario.
Instead, the analysis above shows that for $t_{2g}$ orbitals as in iridium, SOC couples spin component $S^z$ to the fcc lattice bonds lying normal to $\hat{z}$. The simple cubic lattice version of \HKH cannot be generated, and a compound structured as layers of a square lattice would collapse its Kitaev exchange to uniform ising couplings along all square lattice bonds.

\begin{figure}[htb]
\includegraphics[width=82 mm]{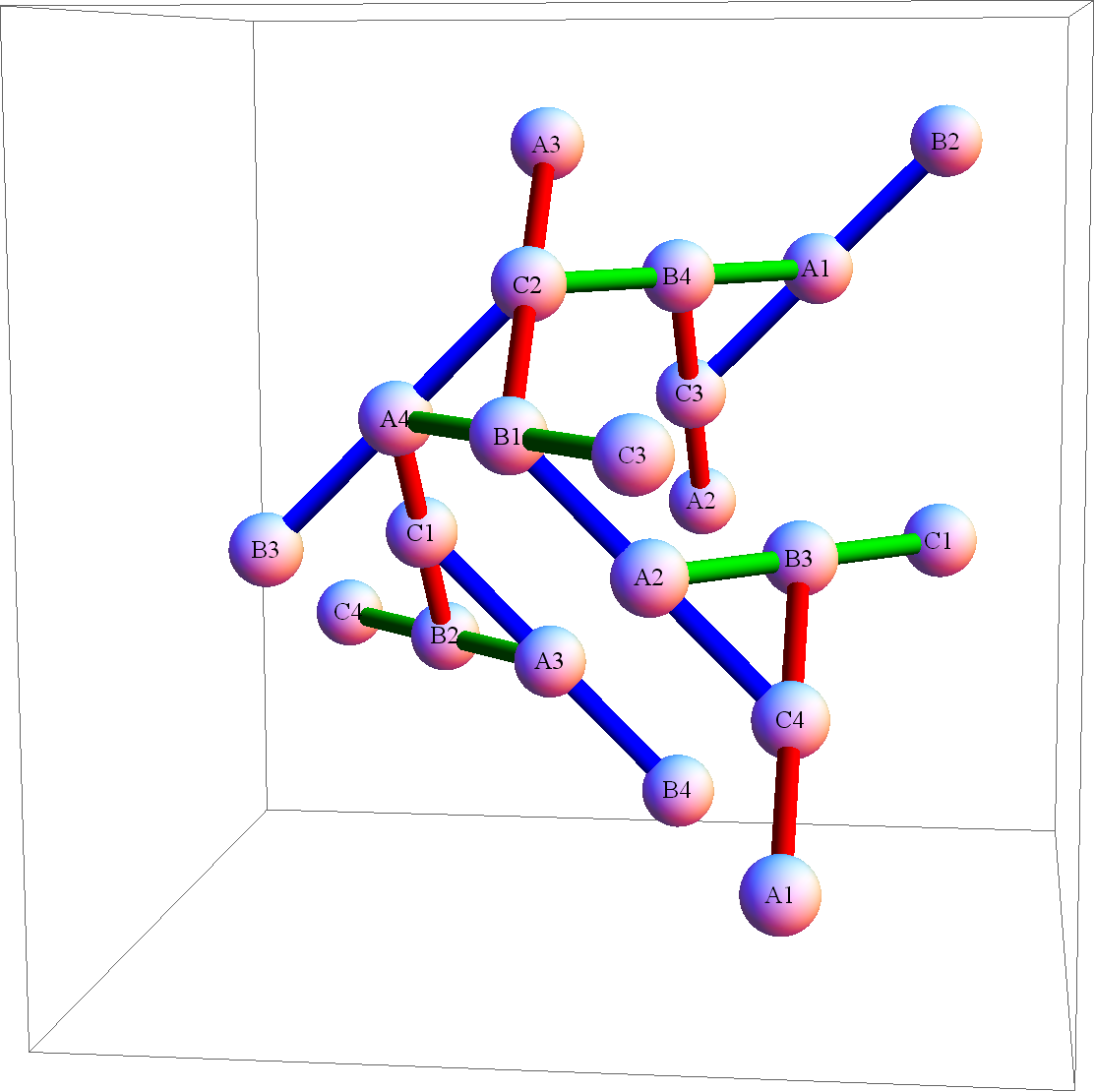}
\caption{{\bf Hyperkagome unit cell and decomposition into Kitaev clusters. }
A symmetric depiction of the hyperkagome structure, highlighting the four disjointed three-site clusters which split up the unit cell for a given Kitaev bond label (red,green,blue).
The fact that the lattice fragments into these disjointed clusters for a given Kitaev bond type has substantial repercussions, as described in the text.
The four clusters of a given type appear in two parallel-orientation pairs which are perpendicular to each other (with different shading). The unit cell is composed of the 12 sites participating in the four drawn triangular faces, as well as all 24 drawn bonds. We label the 12 sites by a letter A,B,C as in Ref.~\onlinecite{Vishwanath2008} so that A spins lie on midpoints of type-A (here blue) clusters, etc; and by a number 1--4 (chosen to not repeat within a triangle face or cluster).
The camera angle, i.e. the vector pointing into the page, is just slightly off (up-right) of a Cartesian (here also Bravais) axis; for Na$_4$Ir$_3$O$_8$ it is the vector from an iridium ion to a neighboring coordinating oxygen.
}
\label{fig:hyperkagome-unit-cell}
\end{figure}

The honeycomb and hyperkagome lattices share a common feature distinguishing them from the other lattices: if we only keep bonds of a single Kitaev type $\gamma$, the lattice fragments into localized disconnected \textit{clusters}. On the honeycomb, each cluster contains two sites, and forms the unit cell. On the hyperkagome, each cluster contains three sites, arranged into a line segment. For a given bond label $\gamma$, the twelve site unit cell fragments into four disjointed clusters, whose line segments are oriented parallel within each of two pairs and perpendicular between the pairs. The structure on the hyperkagome unit cell is shown in Fig.~\ref{fig:hyperkagome-unit-cell}.
As discussed below, this fact has dramatic repercussions for the Kitaev Hamiltonians in both the Luttinger-Tisza approximation and in the Schwinger fermion Majorana mean field (which on the honeycomb describes the Kitaev QSL). In both cases, certain excitations only propagate within a single Kitaev bond type $\gamma$, and the localized disconnected clusters imply these excitations must have completely flat bands.

\section{Klein duality and hidden ferromagnets}
\subsection{Connections to previous work}
Exactly solvable quantum Hamiltonians are rare in dimension higher than one.
It is quite remarkable that the stripy phase at $\eta=+1,\ \alpha=1/2$ found for the honeycomb Kitaev-Heisenberg model\cite{Khaliullin2010} is exact, a hidden ferromagnet exposed by the site dependent spin rotation which quadruples the unit cell\footnote{Spin exchanges with lower symmetry may even be mapped to antisymmetric Dzyaloshinskii-Moriya exchanges.}. Unlike Neel order on even bipartite lattices, this stripy antiferromagnetic order is exact and fluctuation-free at $\alpha=1/2$.

This "four sublattice rotation trick" has been known by Khaliullin and Okamoto for $t_{2g}$ orbitals in a cubic environment since as early as 2002\cite{Okamoto2002}.
It was used for Kitaev-Heisenberg-like Hamiltonians in ferromagnetic titanates\cite{Okamoto2002,Okamoto2003} as well as in other systems, including an explicit transformation on the triangular lattice\cite{Khaliullin2005}
\footnote{In this work\cite{Khaliullin2005}, see especially the discussions related to Fig. 1 (p.170) and Fig. 5 (p.194).}
 to find the dual of 120$^\circ$ order for CoO$_2$.
It was then applied to the honeycomb lattice by Chaloupka, Jackeli and Khaliullin in their derivation of the Kitaev-Heisenberg model for the honeycomb iridates\cite{Khaliullin2010}.
However, its general structure has not been previously elucidated. We will now show that this duality transformation may be defined on general graphs with Kitaev $\gamma$ bond labels and that it has the structure of the Klein four-group, isomorphic to $\mathbb{Z}_2 \times \mathbb{Z}_2$.
This will then lead to a geometrical condition specifying which lattices and finite graphs admit the Klein duality, a result especially useful for designing finite graphs for numerical studies.

\subsection{Deriving the Klein transformation on graphs with Kitaev bond labels}
We begin by defining a general unitary transformation, and then we will show that under certain special conditions it acts as a duality transformation on  Eq.~\ref{eq:KH}. Throughout this paper, by a ``duality transformation'' we refer to a mapping between Hamiltonians that maps a set of Hamiltonians (and the associated phase diagram) to itself (of course without mapping each particular Hamiltonian to itself). 
Consider a lattice or finite graph in any dimension which connects $S=1/2$ spins, and assume each bond $(i,j)$ carries a Kitaev \textit{type} label
\beq
\gamma_{i,j}\in \{\mathbb{1},x,y,z\}.
\eeq
The set $\gamma \in \{x,y,z\}$ corresponds to Kitaev coupling $S_i^\gamma S_j^\gamma$ on that bond, where $\{x,y,z\}$ identifies a set of orthogonal axes in the spin Bloch sphere. The Hamiltonian on the bond may have other terms such as Heisenberg coupling and various anisotropies; but the transformation will turn out to be most useful if the coupling includes only Kitaev and Heisenberg terms, as in Eq.~\ref{eq:KH}. The label $\gamma_{i,j} = \mathbb{1}$ can be assigned to a bond that does not have a Kitaev exchange (only Heisenberg and possible anisotropies), such as a second or third neighbor interaction.  In general such farther neighbor interactions supplementing  \HKH will frustrate the transformation, so when making use of the Klein duality the lattice should usually be considered to be just the pure nearest neighbor Kitaev-Heisenberg model \HKH, where all bonds carry $\gamma \in \{x,y,z\}$. But we will show below that certain farther neighbor Heisenberg interactions do preserve the duality structure, and may be fruitfully included as $\gamma = \mathbb{1}$.

Let us proceed by describing the relevant transformations on individual sites. Assign each site a label
\beq
a_i\in \{\mathbb{1},X,Y,Z\}
\eeq
 which will specify a unitary transformation on that site, specifically rotation by $\pi$ around the Bloch sphere axis $S^a$ for $a \in \{ X,Y,Z\}$, and no rotation for the identity element $a=\mathbb{1}$. Note that $\pi$ rotation around $S^a$ flips the sign of the spin components perpendicular to $a$, so that the rotation $a_i=Z$ multiplies the $(x,y,z)$ components of $S_i$ by the sign structure $g[Z]=(-1,-1,1)$, and also that $g[\mathbb{1}]=(1,1,1)$.

Now, observe that both bond labels $\gamma_{i,j}$ and site labels $a_i$ may be interpreted as elements of the single set $\{\mathbb{1},X,Y,Z\}$.  We may turn this set into a group by defining a multiplication rule. A possible definition is suggested by the multiplication of the associated sign structures $g$, which entails for example $g[X]g[Y]=g[Z]$, suggesting we should define $X\ Y=Z$. The resulting multiplication table is defined by
\beq
X^2=Y^2=Z^2=XYZ =\mathbb{1}
\eeq
with $\mathbb{1}$ acting as the identity. This is the presentation of the group with generators $(X,Y,Z)$ and relations $(X^2,Y^2,Z^2,XYZ)$, known as the \textit{Klein four-group}. The Klein group is abelian and with four elements is the smallest non-cyclic group; it is isomorphic to $\mathbb{Z}_2 \times \mathbb{Z}_2$.

There is an alternative, \textit{geometrical,} way to define multiplication on the elements $a_i$ and $\gamma_{i,j}$. We define the geometric multiplication $(\ast)$ of a site $i$ and one of its bonds $(i,j)$ to be the site reached by traversing the bond, $i \ast (i,j) = j$.
The associated Klein group elements $a_i$ and $\gamma_{i,j}$ inherit this geometric multiplication as
\beq
a_i \ast \gamma_{i,j} = a_j.
\eeq
The Klein group product ($\times$) and the geometric multiplication $(\ast)$  are consistent on a bond if they give the same answer, $a_i \ast \gamma_{i,j} = a_i \times \gamma_{i,j}$.
We say the transformation given by site labels $\{a_i\}$ is the \textit{Klein} transformation if the geometrical multiplication is consistent with Klein group multiplication on \textit{every} bond in the lattice.

If the transformation site labels $a_i$, $a_j$ across a bond are consistent with the Klein group product, i.e.
\beq
a_i \times \gamma_{i,j} = a_j 
\eeq
or equivalently (since elements in the Klein group square to the identity)
\beq
\quad a_i \times a_j = \gamma_{i,j}\ ,
\eeq
 then the transformation changes the form of a Kitaev-Heisenberg coupling in an especially simple way. This is simply because the sign flips $g[a]$ multiply by the Klein group rules, so the diagonal spin exchange $\sum_\alpha J^\alpha_{i,j}S^\alpha_i S^\alpha_j$ transforms by
 \beq
 J^\alpha_{i,j}\rightarrow g[a_i]_\alpha g[a_j]_\alpha J^\alpha_{i,j} = g[\gamma_{i,j}]_\alpha J^\alpha_{i,j}
 \eeq
where $g[a]_\alpha \in \pm 1$ is component number $\alpha$ of the vector $g[a]$ of $\pm 1$ signs. The transformation flips the sign of the components of $J$ perpendicular to the bond type label. 
For Kitaev Heisenberg exchange, this means that the Heisenberg coefficient flips sign and the Kitaev coefficient gains twice the (old) Heisenberg coefficient.

Even if the transformation labels on two sites are consistent with the Klein group product on that bond, it might seem improbable that the $a_i$ rotation labels can be chosen across the entire lattice in a pattern that is Klein group consistent on all bonds. Such consistency for all bonds is necessary for the transformation to change the Hamiltonian uniformly. Now the Klein group structure shows its worth. The condition on the transformation $\{a_i\}$  --- consistency between geometric and Klein group multiplication on each bond --- can be expressed as a condition which refers only to the lattice: that the $\gamma_{i,j}$ encountered in any closed path multiply to the identity $\mathbb{1}$. In other words, all closed loops on the lattice must be composed of the identity operators $\mathbb{1},X^2,Y^2,Z^2,XYZ$. Then the transformation may be consistently defined by Klein group multiplication of bond labels on a any path,
\beq
a_j = \left(\prod_{\ell \in \text{path}_{i\rightarrow j}} \gamma_\ell \right) a_i .
\eeq

\subsection{Geometrical condition for the Klein duality}
\begin{figure*}[htp]
\includegraphics[width=1.8\columnwidth]{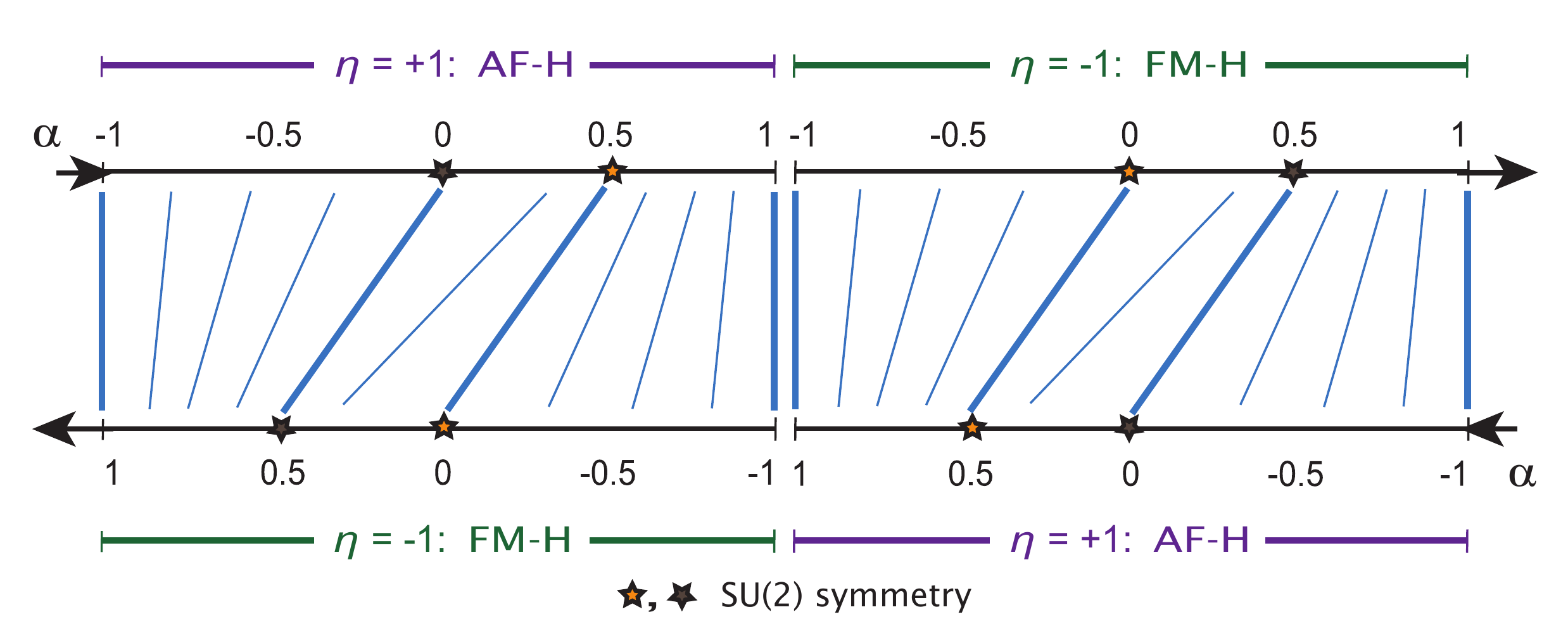}
\caption{{\bf Action of the Klein duality on the Kitaev-Heisenberg Hamiltonian. }
The $\eta,\alpha$ parameters on the top line (oriented left to right) of the Hamiltonian \eqh  map according to the blue lines to $\eta,\alpha$ parameters on the bottom line (oriented right to left) in the Klein-dual Hamiltonian. Thick blue lines map the points shown exactly, thin blue lines are a qualitative sketch. Note that the right and left edges of the figure are identified, forming a ring.
Both pure Kitaev Hamiltonians ($\alpha=\pm 1$) are \textit{self-dual}, mapping to themselves. The points at $\alpha=1/2$ and $\eta=+1$, $\eta=-1$ are dual to the SU(2) symmetric FM-Heisenberg and AF-Heisenberg points (yellow and brown stars) respectively, with the FM dual point hosting the exactly soluble \textit{stripy} phases on all lattices in Fig.~\ref{fig:KH-lattices}.
\label{fig:Kleinduality}}
\end{figure*}

We have shown that the existence of the Klein duality can be expressed as a condition on the lattice. Using the Klein group structure, we can write this condition as follows. \textit{Any closed loop, containing $N_x$ $x$-bonds, $N_y$ $y$-bonds and $N_z$ $z$-bonds, must satisfy}
\beq
N_x,\ N_y,\ N_z\quad \text{all even or all odd.}
\label{eq:Klein-constraint}
\eeq
The three $N_i$s can be all even because Klein group elements square to the identity, or all odd because $XYZ=1$.
If this condition Eq.~\ref{eq:Klein-constraint}  is satisfied on all closed loops then the Klein duality can be constructed consistently as follows:
choosing a reference site $i$ which for simplicity will be unchanged in the duality,  $a_i = \mathbb{1}$,
assign any site $j$ a rotation label $a_j$ as simply the Klein group product of the Kitaev bond labels $\gamma$ on any path from $i$ to $j$. The constraint Eq.~\ref{eq:Klein-constraint} ensures this duality construction is consistent regardless of the choice of paths $i$ to $j$.
The Klein duality then maps Eq.~\ref{eq:KH} to itself, transforming the parameters $\alpha$ and $\eta$ according to Fig.~\ref{fig:Kleinduality}.

It is easy to see that the Klein duality indeed exists on all of the infinite lattices shown in Fig.~\ref{fig:KH-lattices}; because the Klein group is abelian, it is sufficient to check that the condition is satisfied on small local loops. For example, triangle faces have $N_x=N_y=N_z=1$.  The condition also holds on other lattices such as the simple cubic that can host symmetric Kitaev exchange but cannot generate it via $t_{2g}$-$p$ orbital superexchange. Adding pure Heisenberg ($\gamma=\mathbb{1}$) further neighbor interactions generally spoils the Klein duality, though if all the resulting loops satisfy Eq.~\ref{eq:Klein-constraint}, the Klein duality survives unscathed and moreover does not modify the pure Heisenberg $\gamma=\mathbb{1}$ interactions, even while it flips the sign of Heisenberg interactions on Kitaev-labeled bonds. This occurs, for example, with $J_3$ Heisenberg exchanges on the honeycomb and kagome lattice, connecting sites on opposite corners of a hexagon. The family of Hamiltonians preserved by the duality is then enlarged to $J_K-J_1-J_3$, ie nearest neighbor Kitaev-Heisenberg plus third neighbor Heisenberg. This $J_K-J_1-J_3$ family of Hamiltonians maps to itself (non-trivially) under the Klein transformation, with $J_3$ unchanged. 

For graphs and finite lattices with periodic boundary conditions (PBC),
 considering small local loops is insufficient; loops traversing the PBC may break Eq.~\ref{eq:Klein-constraint} and spoil the Kitaev duality. Such winding loops must be checked explicitly. Here the condition Eq.~\ref{eq:Klein-constraint} should serve much practical use, as finite sized versions of the Fig.~\ref{fig:KH-lattices} lattices with PBC are useful for numerical studies, and it may otherwise be difficult to construct or identify the choice of PBC which admit the Klein duality. In many cases the appropriate PBC involve nontrivial twists that, in a continuum limit, appear as cutting and gluing operations on the boundaries.

\subsection{The Klein duality on $H_\text{KH}$, Klein self-dual points and Klein $\mathbb{Z}_2$ symmetry}
In order to describe the action of the Klein duality on the parameter space of \eqh, let us first discuss the $\eta,\alpha$ snd other parametrizations. In \eqh, the sign $\eta=\pm 1$ is the sign of the Heisenberg exchange and the sign of $\alpha$ is minus the relative sign between the Kitaev and Heisenberg exchanges. This is a compatible extension of the $\alpha$ parametrization introduced in Ref~\onlinecite{Khaliullin2010}; restricting to $\eta=+1$, $0\leq \alpha \leq 1$ gives the original parameter space\cite{Khaliullin2010} with antiferromagnetic (AF) Heisenberg and ferromagnetic (FM) Kitaev interactions. It is clear that both the FM and AF pure Kitaev Hamiltonians are each described by two parameter points, which must be identified,
\beqa
(\eta=+1,\alpha=+1)\ \cong \ (\eta=-1,\alpha=-1) \nonumber \\
(\eta=+1,\alpha=-1)\ \cong \ (\eta=-1,\alpha=+1) .
\eeqa
Identifying (i.e. gluing) these pairs makes the $(\eta,\alpha)$ parameter space into a circle. In the axis shown at the top of Fig.~\ref{fig:Kleinduality}, the two $\alpha$
segments (for $\eta=+1,-1$) are connected both in the middle where they are drawn to almost touch and also at their distant endpoints (where arrows are drawn).
 Comparing to the angular parametrization presented in Ref.~\onlinecite{Khaliullin2012}, $\eta=+1$($\eta=-1$) is the right(left) side of the circle, and $\alpha=-1,...,+1,-1,...,+1$ increases going clockwise.
(We will also sometimes refer to both FM and AF pure Kitaev Hamiltonians simultaneously, in which case the notation   $\alpha=\pm 1$ is unambiguous.) Note that the $\eta,\alpha$ parametrization, though (piecewise) linear, is non-analytic at $\alpha=0,\pm 1$, which may be an issue for certain numerical computations.

Now we may discuss how the Klein duality acts on \eqh.
In other words, the Hamiltonian \eqh  with certain parameters $\eta,\alpha$ is equivalent to the Hamiltonian \eqh on the rotated spins but with different parameters $\eta',\alpha'$. The duality is shown by the blue lines in Fig.~\ref{fig:Kleinduality}. Note that where the blue lines are roughly vertical, the duality approximately just flips the sign of both $\eta$ and $\alpha$, i.e. just flips the sign of the Heisenberg term. In general it flips the sign of the Heisenberg term but also adds twice the (old) Heisenberg term to the Kitaev term,
\beq
J_H \vec{S}_i\cdot\vec{S}_j + J_K S^{\gamma_{ij}}_i S^{\gamma_{ij}}_j \longrightarrow (-J_H) \vec{S}_i\cdot\vec{S}_j 
+ (J_K+2 J_H) S^{\gamma_{ij}}_i S^{\gamma_{ij}}_j .
\label{eq:KHkleinduality}
\eeq
Note that with \eqh as written, changing $\alpha$ also changes the overall energy scale; this can be avoided by dividing \eqh by $(1-|\alpha|)$, so that the magnitude of the Heisenberg term remains fixed at $1$.

On \eqh the duality always takes $\eta\rightarrow -\eta$, but acts on $\alpha$ in a nonlinear way, approximately shown by the changing slope of the blue lines in Fig.~\ref{fig:Kleinduality}. The relation between $\alpha$ and $\alpha'$ is given implicitly by
\beq
\alpha' \alpha = (1-c_{\alpha'})(1-c_\alpha) \ \ , \ \ \
c_\alpha\equiv \left\lbrace
\begin{array}{ll}
0 & \alpha\leq 0 \\
2\alpha & \alpha \geq 0 \end{array} \right.
.
\eeq
As is clear from Fig.~\ref{fig:Kleinduality}, the relation may be written explicitly as a simple piecewise function,
\beq
\alpha'  =  \left\lbrace\begin{array}{ll}
\frac{1}{\alpha+2} & -1\leq\alpha\leq 0 \\
\frac{1-2\alpha}{2-3\alpha} & 0\leq\alpha\leq \frac{1}{2} \\
\frac{1}{\alpha}-2 & \frac{1}{2}\leq\alpha\leq 1
\end{array} \right.
.
\eeq

The family of Hamiltonians \eqh can be generalized by modulating the strength of couplings on different bonds in arbitrary ways; the Klein rotation generalizes as well to arbitrary configurations of coupling strengths. Generically it will no longer map one simple family of Hamiltonians to itself, but it may still offer hidden exactly solvable points, such as by mapping \eqh with toric code anisotropies of the Kitaev coupling strength into a mixed ising-Heisenberg ferromagnet with an exact ground state. Specifically, given Kitaev bond strengths of $(1-a/2,1-a/2,1+a)$ on the three bond types, the location of the hidden ising ferromagnet shifts to $\eta=+1$, $\alpha = 1/(2-a/2)$.

Duality relations in condensed matter physics typically map order to disorder or strong coupling to weak coupling,  such as the duality relating the paramagnetic and ferromagnetic phases in the transverse field (quantum) ising model. The Klein rotation is a duality in the sense of mapping a family of Hamiltonians to itself, but it is not amenable to this typical interpretation.  First, there is no sense of weak and strong coupling regimes within the parameter space of \eqh. Second, this parameter space forms a ring, and rather than a single self-dual point, it offers \textit{two} distinct Hamiltonians which are self-dual under the Klein duality. Third, as is rigorously known for the honeycomb lattice and suggested by the LTA for the other lattices below, the self-dual pure Kitaev Hamiltonians lie in the interior of a phase rather than signifying a phase boundary.

 The Hamiltonians at the two Klein self dual points may alternatively be interpreted as possessing an enlarged symmetry group. The additional symmetry is generated by the Klein duality and has  $\mathbb{Z}_2$ characteristic. It thus acts in a highly nontrivial manner on spins on different sites. Phases that preserve this Klein $\mathbb{Z}_2$  symmetry must contain this highly nontrivial structure; there is currently one known example of such a phase, the Kitaev honeycomb spin liquid.
If any lattice turns out to host a magnetically ordered phase which does not spontaneously break the  Klein $\mathbb{Z}_2$ symmetry, such a phase would have a complex pattern of noncoplanar spin order. This is unlikely, but there may also be phases which break the  Klein $\mathbb{Z}_2$ symmetry but do not break too many other symmetries, yielding a ground state manifold that naturally splits into the two  Klein $\mathbb{Z}_2$ broken portions. Determining which or whether any of these scenarios holds on any particular lattice is left for future work.

\begin{figure*}[tb]
\subfloat[][{\normalsize 3D-stripy order on the face centered cubic (fcc) lattice. } ]{
\includegraphics[width=70 mm]{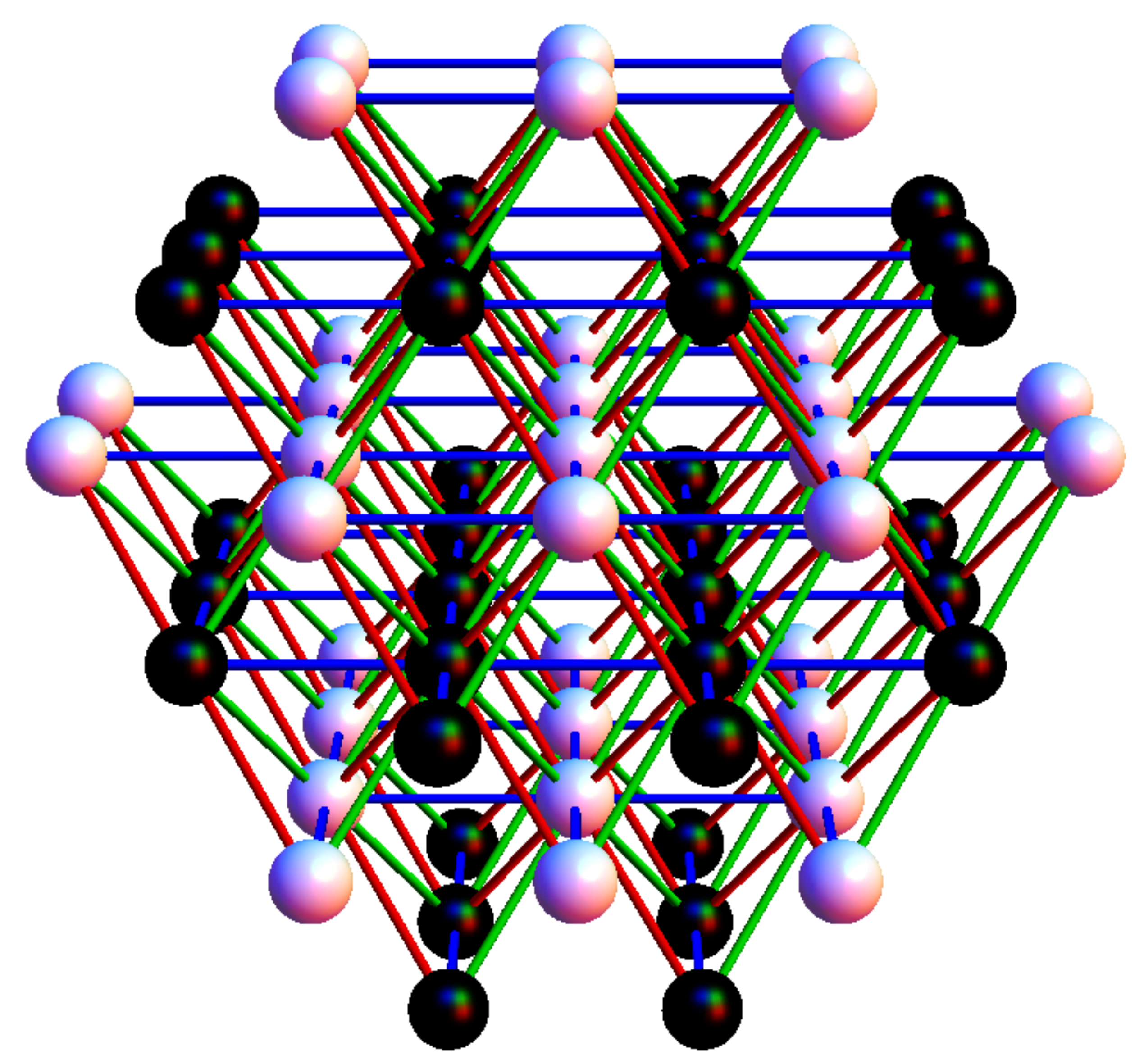}
\label{fig:fcc-stripy}  }
\subfloat[][{\normalsize 3D-stripy order on the pyrochlore lattice.}  ]{
\includegraphics[width=80 mm]{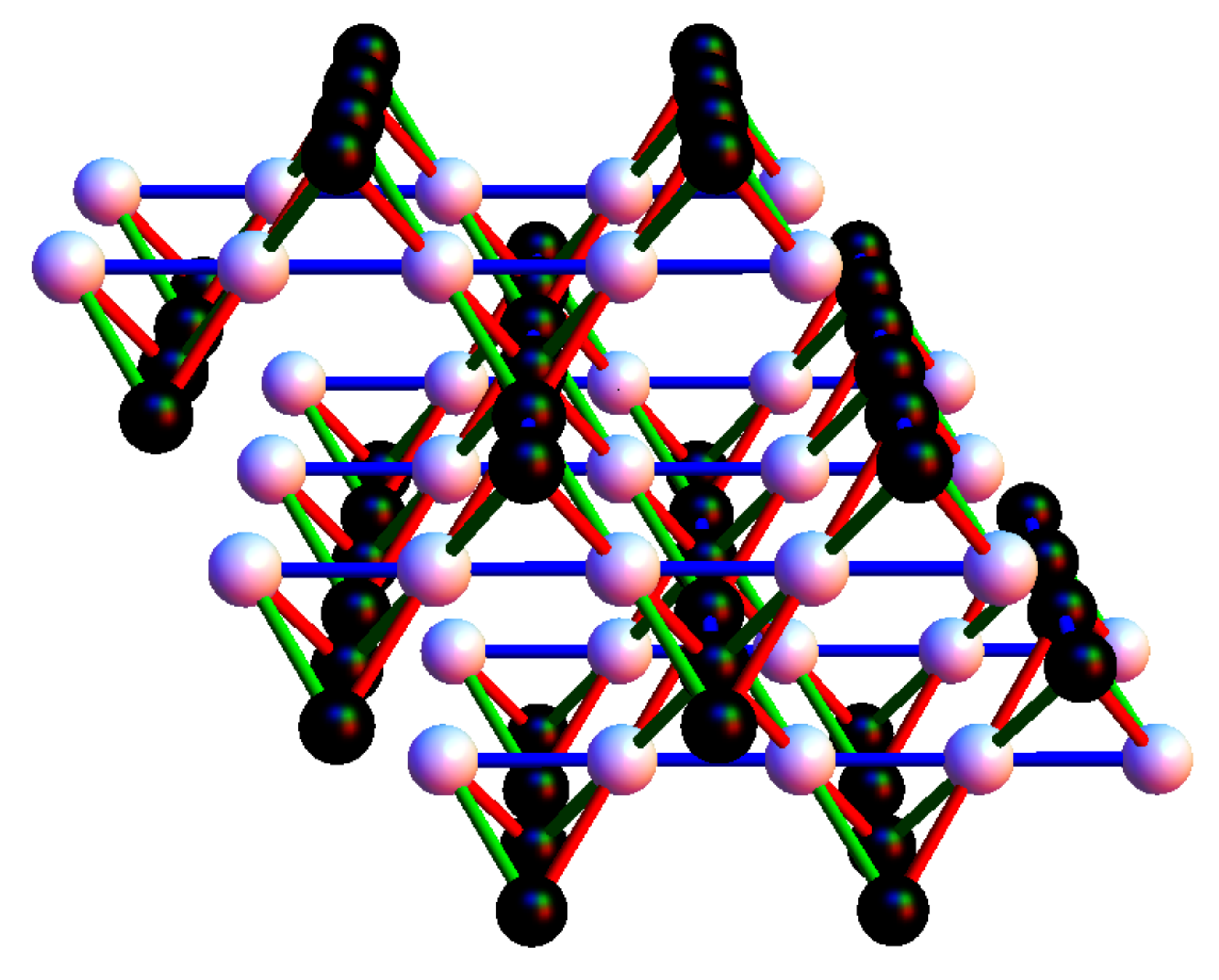}
\label{fig:pyrochlore-stripy} }
\\
\subfloat[][{\normalsize 3D-stripy order on the hyperkagome lattice. } ]{
\includegraphics[width=100 mm]{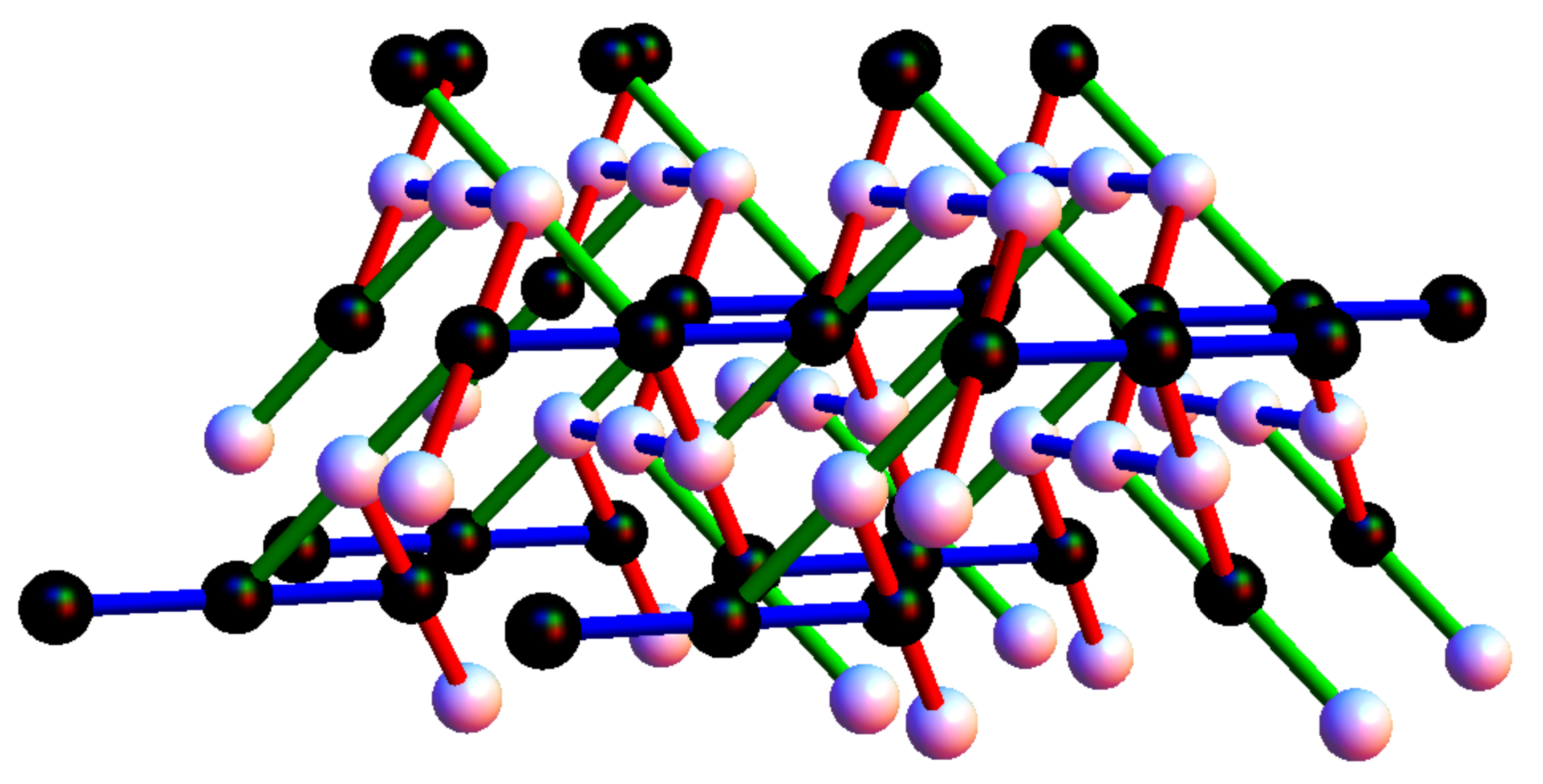}
\label{fig:hyperkagome-stripy} }
\caption{{\bf 3D-Stripy orders on the fcc, pyrochlore and hyperkagome lattices.} Black(white) spheres represent up(down) spins. Bonds are colored red, green, and blue according to the Kitaev label.
The ordering pattern is of alternating planes, here normal to $\hat{z}$; $z$-type bonds (blue), i.e. those within the planes, connect spins of the same orientation.
On the pyrochlore and hyperkagome, planes are broken up into uniformly oriented chains, with chains in spin-up planes oriented perpendicularly to chains on spin-down planes. On the hyperkagome, the chains are further broken into the linear three site clusters shown in Fig.~\ref{fig:hyperkagome-unit-cell}.
These stripy orders are exact at $\eta$=$+1,\alpha$=$1/2$, being Klein-duals of the ferromagnet.
}
\label{fig:all-stripy}
\end{figure*}

 \section{Exactly soluble stripy phases as Klein duals of the ferromagnet}
The most obvious consequence of the existence of the Klein duality is seen by applying the duality on the Heisenberg ferromagnet. At the resulting parameter point $\eta=+1$, $\alpha=1/2$, the ground state manifold of the quantum Hamiltonian is known exactly and consists of simple product states, parametrized by the full SU(2) symmetry. The ground states may be found by taking a ground state of the Heisenberg ferromagnet, and applying the rotations defined by the Klein duality on this magnetic order. The result is the \textit{stripy} collinear magnetic order. We will use the name \textit{stripy} to refer to the FM-dual phase on lattices in any dimension, both to preserve the analogy to the honeycomb and also because, as shown below, the 3D-stripy orders can have some ``stripy'' features in their own right.

Away from the SU(2) symmetric point the symmetries reduce to the lattice SOC operations. The stripy ordering breaks the three-fold rotation symmetry, present in all Kitaev lattices as in Fig.~\ref{fig:KH-lattices}, that simultaneously permutes the suitably chosen Euclidean directions $\hat{x}\rightarrow\hat{y}\rightarrow\hat{z}\rightarrow\hat{x}$, the same axes on the Bloch sphere and also the Kitaev bond labels $x\rightarrow y\rightarrow z\rightarrow x$.
The appropriate coordinate system is set by an IrO$_6$ octahedron, in which the ordering is along one of the three directions $(1,0,0),(0,1,0),(0,0,1)$ i.e. $\hat{x},\hat{y},\hat{z}$. $\hat{z}$-type stripy order has $z$-bonded spins aligned parallel and $x$ or $y$ bonded spins aligned antiparallel. The collinear spin axis is then fixed to $S^z$, though the direction of the ordered moment will likely be determined by other effects in any material realization.

The stripy orders on the various lattices share common features but also host distinguishing characteristics.
On the two dimensional lattices, which always appear as layers perpendicular to the $(1,1,1)$ axis in the IrO$_6$ coordinate system, the ordering breaks the (SOC version of) $120^\circ$ lattice rotation symmetry. On the triangular lattice it is literally alternating stripes (i.e. lines of sites) of up spins and down spins. On the honeycomb lattice, each stripe is composed of the two-site clusters that lie on a given line; this order is also known as ``IV'' in the $J_1$-$J_2$-$J_3$ literature. On the kagome lattice, stripy order gives the same configuration on each unit cell (is wavevector $\Gamma$) of two spins up and one spin down, meaning it is \textit{ferrimagnetic} with a nonzero net magnetization. At the exact $\alpha=1/2$ point the spins are saturated and the net magnetization is $1/3$ that of the ferromagnet.

In three dimensions, the 3D-stripy orders involves alternating planes of up spins and down spins. For say $\hat{z}$ stripy order the planes are normal to $\hat{z}$.
On the face centered cubic (fcc) lattice, the planes are faces of the fcc cube.
On the pyrochlore lattice, the stripy order acquires an additional feature: spin-up planes are broken up into \textit{chains} aligned in one particular direction, and spin-down planes are composed of chains aligned in the perpendicular direction. On the hyperkagome lattice this feature persists, and moreover the chains are broken into oriented linear clusters: for $z$-stripy order the $z$-type three-spin-chain clusters of Fig.~\ref{fig:hyperkagome-unit-cell} are oriented uniformly within the spin-up planes, and also uniformly but in a perpendicular orientation within the spin-down planes. The 3D-stripy orders are shown in Fig.~\ref{fig:all-stripy}.

\section{Luttinger-Tisza approximation phase diagrams}
Except for the Heisenberg ferromagnet and its Klein dual point as described above,
the Kitaev-Heisenberg Hamiltonians are \textit{frustrated}
\footnote{the antiferromagnet and its Klein dual on the bipartite honeycomb are of course exceptions as well.}.
The resulting sign problem for quantum Monte Carlo renders their quantum phase diagrams, especially for the three dimensional lattices, exceedingly difficult to compute.  The Hamiltonians Eq.~\ref{eq:KH} on the various lattices are quite unique in that they all offer an exact solution at a nontrivial point in the phase diagram, the Klein dual to the ferromagnet.  To explore the remainder of the phase diagrams we must use  approximation methods, as we shall now describe.

For an initial survey of the phase diagrams we employ the Luttinger-Tisza Approximation (LTA), also known as the spherical model\cite{Tisza1946,Luttinger1951,Anderson1950,Menyuk2007}. It is a semiclassical approximation in that it improves upon the classical Hamiltonian, incorporating some notion of quantum fluctuations and a reduced ordered moment. While the classical version of a Hamiltonian has the hard constraint that the ordered moment (i.e. the spin vector) on each site must have magnitude $S$, quantum fluctuations are expected to relax this constraint.  
Implementing this constraint only on average with a single global Lagrange multiplier, the Hamiltonian \eqh becomes free quadratic and the lowest energy configuration of the classical spins may simply be found by a Fourier Transform and a diagonalization of the spin and sublattice indices.

The LTA always computes a lower bound to the energy of the \textit{classical} model;
this inequality becomes a strict equality when the LTA minimum energy configuration happens to obey the unit length constraint.
 In turn, classical configurations of spins with length $S$  give upper bounds to the true  ground state energy of a spin-$S$ quantum Hamiltonian\cite{Anderson1951}, simply by defining site-product wave functions which by the variational principle have at least the ground state energy.
When a non-normalized configuration is chosen by the LTA, its energy is lower than the classical minimum energy which in turn is generally higher than the quantum ground state energy, so the energy of the LTA configuration can match the true ground state energy. Relaxing the unit length constraint indeed allows the classical ordered moments to fluctuate, and in some ways improves upon the constrained classical Hamiltonian as an approximation to the quantum Hamiltonian.

On a Bravais lattice and for the case with SU(2) spin rotation symmetry, solutions with normalized spins can always be constructed from the LTA minimum eigenvalues\cite{Menyuk2007}. For momenta $q$ satisfying $q=-q$ there is a family of degenerate orders but even for arbitrary incommensurate momenta there are coplanar spiral solutions with normalized spins, with the first(second) spin component modulated by the real(imaginary) part of $\exp[i q r]$. However, when SU(2) spin rotation symmetry is broken such as by SOC, there may only be one low energy spin component and this approach can fail, requiring $q=-q$ to construct states with unit length normalized spins. On lattices with multiple sites per unit cell, the LTA may assign different lengths to sites in the unit cell, which again points to frustration, though if the spins have nearly the same length then we expect that ordering pattern to be robust\cite{Henley2012}. Note that even when classical solution do exist, when the LTA identifies extensive ground state degeneracy or includes degenerate ground state configurations with vanishing ordered moment, it suggests quantum fluctuations will melt any magnetic order. In such cases determining the ground state requires a full quantum analysis.  Thus while the Luttinger-Tisza approximation cannot characterize non-classical phases, it is a useful first approach for identifying features in the phase diagram.

The LTA phase diagrams are shown in Fig.~\ref{fig:O3-diagrams}. Here we discuss general features; see the Appendix for details. Stripy phases are found surrounding the FM-dual point in all of the lattices; they are exact ground states at $\eta$=$+1,\alpha$=$1/2$ even within the LTA. However, the kagome and hyperkagome lattices exhibit an interesting frustration: while spins are uniformly normalized at the SU(2) FM-dual point, away from $\alpha=1/2$ the energy is minimized when spins within the unit cell are of different lengths. As it must by the Klein duality, this frustration is observed in the ferromagnet phase as well. Evidently for the kagome and hyperkagome, but not for the pyrochlore or the other lattices, even small SU(2) breaking within the ferromagnetic phase creates substantial frustration visible in the LTA.

At certain points in the phase diagram, all wavevectors in the BZ offer spin configurations with the same minimum energy, so that the lowest band is flat. While subextensive degeneracies occur generically at certain parameter points and are expected to be completely lifted by boundary conditions, such extensive degeneracies, marked by ``$Q$'' in Fig.~\ref{fig:O3-diagrams}, likely signify a new phase. What could the new phase be? There are only two Hamiltonians hosting LTA extensive degeneracies for which the quantum ground state is known: the honeycomb Kitaev model  ($\alpha=\pm 1$) which is exactly soluble, hosting the Kitaev QSL with Majorana fermionic spinons;  and the kagome Heisenberg antiferromagnet, which was recently found by DMRG simulations\cite{White2011, Balents2012} to host a QSL phase, consistent with a bosonic $\mathbb{Z}_2$ QSL\cite{Sachdev1992}. The ground states of pyrochlore and hyperkagome Heisenberg antiferromagnets, which also have LTA flat bands, are not conclusively known but have been proposed to be plaquette or dimer valence bond solids (VBS) as well as various fractionalized QSLs\cite{Moessner2010b,Balents2008b,Zhang2008b,Kawakami2001}.

There are thus two conclusions to draw about the other $Q$ points in Fig.~\ref{fig:O3-diagrams}. First, by the Klein duality, any lattice hosting a phase with no magnetic order in its AF Heisenberg model also has the same type of phase surrounding the $\eta$=$-1,\alpha$=$1/2$ point, with FM Heisenberg and AF Kitaev exchanges. For example, the recent discovery of the kagome AF Heisenberg QSL then immediately yields the Klein dual of this QSL at the dual point; this Klein dual QSL will likely have distinct physical properties in its response to external fields.
Second, by analogy with the known $Q$ points mentioned above, we may guess that the pure Kitaev models on the hyperkagome also host a quantum phase with no magnetic order, either a VBS or a QSL.

It is especially encouraging that the LTA flat bands within the honeycomb and the hyperkagome pure Kitaev models arise via the same mechanism.
Consider that LTA flat bands in the AF Heisenberg models occur due to the lattice specific band structure from a hopping model with $\pi$ flux.
For the pure Kitaev models $\alpha = \pm 1$, a given spin component such as $S^z$ hops only on $z$-type bonds. As mentioned above, for the honeycomb and hyperkagome lattices, turning off $y$ and $x$ bonds splits the lattice into an extensive number of localized disconnected segments, as shown for the hyperkagome in Fig.~\ref{fig:hyperkagome-unit-cell}. Localization in the disjointed clusters yields the flat bands. Moreover, unlike for the Heisenberg case where (in the relevant lattices we study) there are gapless excitations where the flat lowest band touches higher ones, for the Kitaev cases the disjointed clusters yield a band structure where all bands are completely flat and fully gapped, in the hyperkagome case also fourfold degenerate at each wavevector due to four clusters in the unit cell.

Returning to the survey of the LTA phase diagrams, we find other regions with strong frustrations. On the kagome and pyrochlore lattices, over wide regions of parameter space the LTA fails spectacularly: in the regimes labeled ``frustrated'', the unit cell in both lattices has two spins aligned antiparallel but with the remaining one (kagome) or two (pyrochlore) spins chosen to have exactly zero ordered moment by the LTA. Viewing the LTA as an enhancement of classical solutions which incorporates quantum fluctuations, we see that here the expected quantum fluctuations are sufficiently strong to eliminate some of the ordered moments, pointing to especially strong quantum frustration.   A related regime on the fcc, a Bravais lattice, finds subextensive degeneracy involving incommensurate momenta, that would form spiral orders but with only one low energy spin component cannot achieve correctly normalized spins across the spiral.

\begin{figure}[ptb]
\includegraphics[width=80 mm]{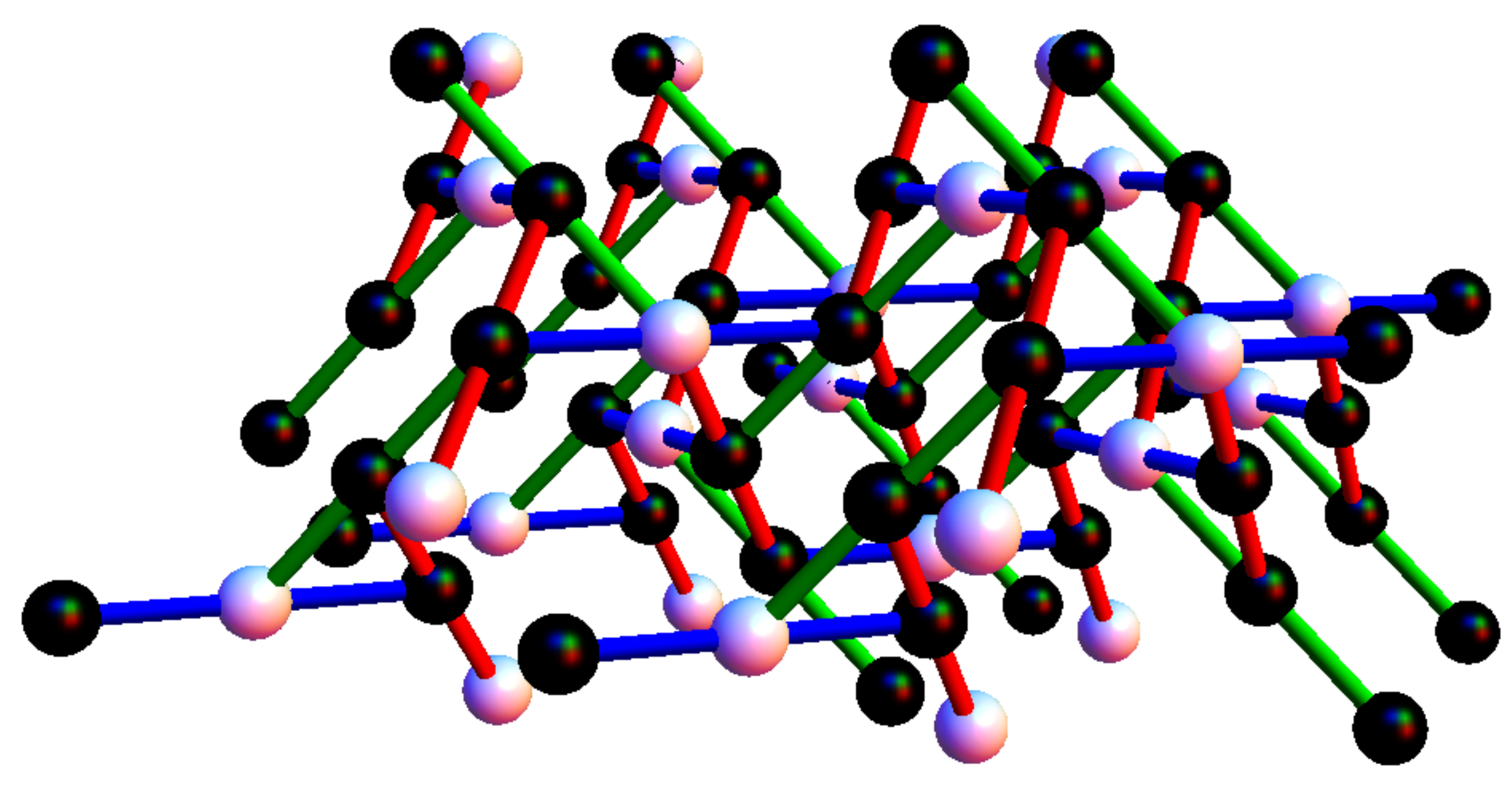}
\caption{{\bf Magentic order of the cluster-ferrimagnet state on the hyperkagome.}
Black(white) spheres are up(down) spins; bonds are colored according to Kitaev label.
Here shown for $\hat{z}$ ordering, notice that the $z$-type clusters (blue), lying in planes normal to $\hat{z}$, all have ``(up, down, up)'' spin configurations. This configuration has \textit{nonzero} net magnetization. Note that the cluster-ferrimagnet order is Klein-dual to the cluster-AF order. 
  }
\label{fig:hyperkagome-clusterferri}
\end{figure}

\begin{figure}[ptb]
\includegraphics[width=80 mm]{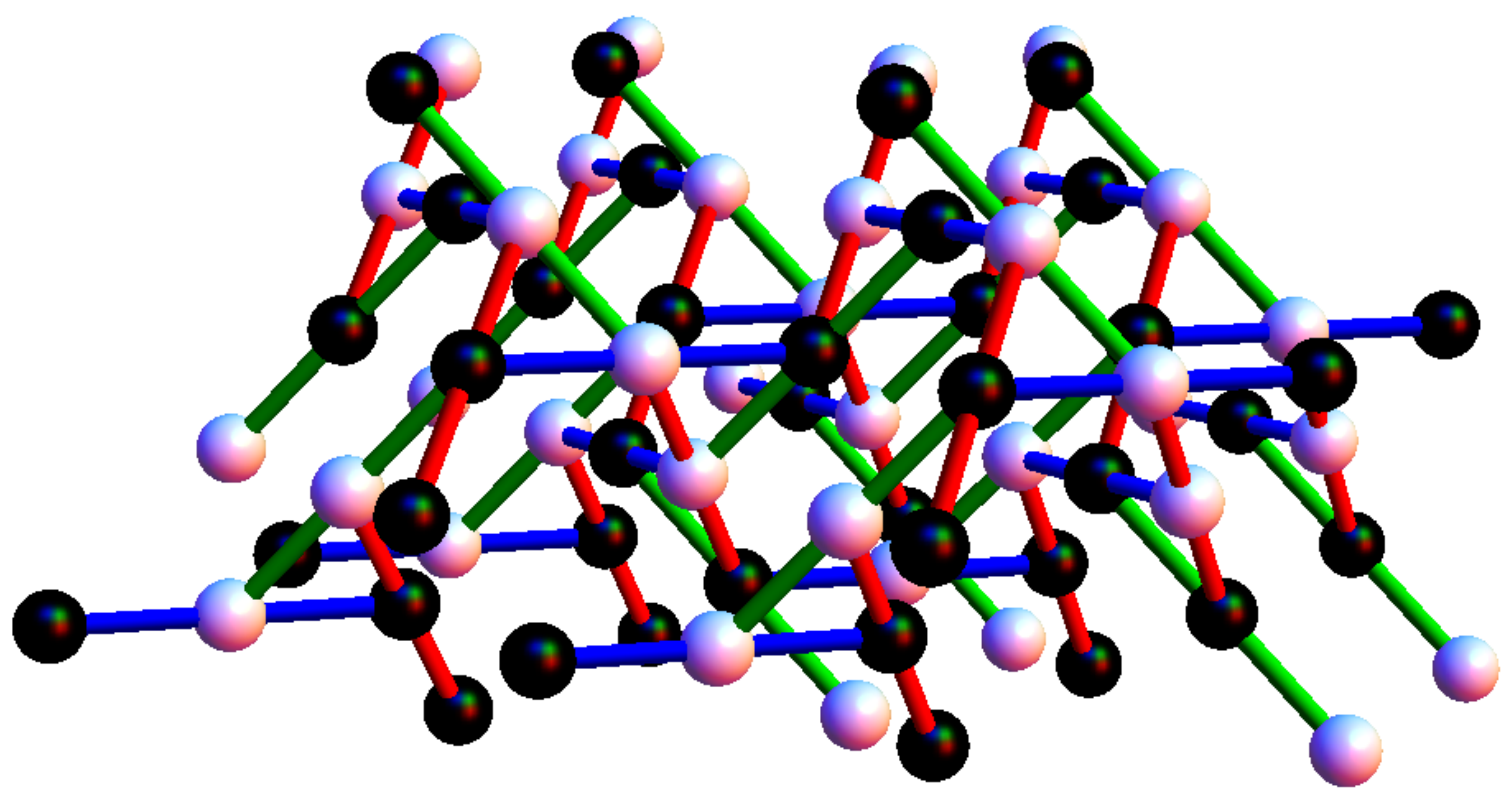}
\caption{{\bf Magentic order of the cluster-AF state on the hyperkagome.}
Black(white) spheres are up(down) spins; bonds are colored according to Kitaev label.
Here shown for $\hat{z}$ ordering, notice that the $z$-type clusters (blue), lying in planes normal to $\hat{z}$, have ``(up, down, up)'' spin configurations on even-numbered planes, and the opposite ``(down, up, down)'' spin configurations on odd-numbered planes. This configuration has \textit{zero} net magnetization. Note that the cluster-AF order is Klein-dual to the cluster-ferrimagnet order. 
  }
\label{fig:hyperkagome-clusterantiferro}
\end{figure}

Finally, on the hyperkagome the LTA finds two regimes with apparent magnetic order with unconventional spin configurations. Though in both cases the spins cross the unit cell are not chosen to have the same ordered moment, this is expected with such a large unit cell, and the LTA configurations should serve as good starting points for quantum Hamiltonian ground states, likely with quantum fluctuations greatly reducing the ordered moment. For $\eta$=$+1$ and $\alpha<0$ we find that for $z$-type order, $z$-clusters all have the identical spin ordering ``(up, down, up)'', resulting in an AF state with a nonzero net magnetization which we thus term the ``cluster-ferrimagnet''. The Klein dual of this order, for large $\alpha$ at $\eta$=$-1$, has the same ``(up, down, up)'' pattern in each $z$-cluster except clusters are flipped on alternating planes, so there is zero net magnetization; we term it the ``cluster-AF'' state. These two Klein dual orders are shown in Fig.~\ref{fig:hyperkagome-clusterferri} and Fig.~\ref{fig:hyperkagome-clusterantiferro}.

\section{Searching for analogues of the Kitaev Majorana spin liquid beyond the honeycomb}
All the lattices in Figure \ref{fig:KH-lattices} except the honeycomb have coordination number larger than three, spoiling the Kitaev honeycomb spin liquid exact solution. However, similar Majorana QSL phases could still occur for the Kitaev Hamiltonians on the other lattices,
only without an exact solution and with nonzero correlation length.  Since it is generally highly difficult to determine whether the true ground state of a spin Hamiltonian forms a QSL, we will not attempt to answer this question. Instead, we will study possibilities for similar Majorana QSLs on the other lattices using an appropriate choice of mean field.

The exact solution of the Kitaev honeycomb model in terms of Majorana fermion operators is a specific case of a Schwinger fermion decomposition mean field, which becomes \textit{exact} for this model\cite{Vishwanath2012,Nayak2011,Okamoto2013}.
To search for similar Majorana QSLs on the other lattices, we thus employ this mean field. Spins are decomposed into bilinears in four Majorana species $\chi^{0,1,2,3}$ as
\beq
 S^a \rightarrow i \chi^0 \chi^a \ \ \ \text{with}\ \ \{ \chi^a,\chi^b \}=\delta^{a,b} .
\eeq
This mapping is exact under the single fermion occupancy constraint $\chi_{i}^{0}\chi_{i}^{1}\chi_{i}^{2}\chi_{i}^{3}=1/4$. On the honeycomb lattice this constraint commutes with the pure Kitaev Hamiltonian, but that does not occur on the other lattices. The $\mathbb{Z}_2$ gauge freedom in defining the Majorana operators enables a choice of attaching gauge transformations to the physical symmetry operations, called a projective symmetry group (PSG)\cite{Wen2002}; see the Appendix for details. The PSG of the Kitaev honeycomb model was previously studied\cite{Vishwanath2012} and determined to be flux-free, with $(\chi_{1},\chi_{2},\chi_{3})$ transforming as a pseudovector and each bond permitting Majorana bilinear expectation values only for two Majoranas of the same species $a$, yielding a total of three mean field parameters:
\beq
u_{\gamma}^{0}\equiv u_0,\ u_{a}^{a}\equiv u_a,\ u_{a}^{b\neq a}\equiv u_b
\eeq
with
\beq
 u_{\gamma[v]}^{\alpha}\equiv\langle i\chi_{j}^{\alpha}\chi_{j+v}^{\alpha}\rangle
\eeq
where $b$ is a bond.  $u_b$ is set to zero for the pure Kitaev model.
The resulting mean field Hamiltonian is
\bea
&H_{\text{MF}}=-\frac{1}{2}\sum_{i,v[i],\alpha}\text{sign}[v]\nu_{\gamma[v]}^{\alpha}i\chi_{i}^{\alpha}\chi_{j}^{\alpha}
\nonumber \\
&  \nu_{\gamma}^{a}=\eta\left((1-|\alpha|)-2\alpha\delta_{\gamma}^{a}\right) u_0 \nonumber \\
& \nu_{\gamma}^{0}=\sum_a J^a_\gamma u^a_\gamma= \eta\left((2-2\alpha-2|\alpha|)u_b -2\alpha u_a \right)
\eea
where $J^a_\gamma$ is the coupling of spin component $a$ on a $\gamma$-bond, $i$ is a site, $v[i]$ are the bonds of site $i$ and sign$[v]$ is the orientation of the bond $v$ within the PSG. This orientation determines how operators on the bond transform under symmetries. On the honeycomb, bonds are oriented from sublattice A to B.
The bond orientations used in the PSGs for the triangular and the kagome lattice are depicted in Ref.~\onlinecite{Sachdev1992} (though that work dealt with bosonic QSLs, the bond orientation diagrams we take are the same). For the triangle, it is known as the zero flux PSG. For the kagome, this zero flux PSG is known as $\sqrt{3}\times\sqrt{3}$ or $Q_1 = -Q_2$. The PSG analysis for this type of mean field has not been successfully carried out on the 3D lattices; the pyrochlore does not appear to give a unique decomposition\cite{Sondhi2006}.
On the hyperkagome however one of the four spins in each tetrahedra is removed, so we can consistently choose the orientation  $A\rightarrow B\rightarrow C\rightarrow A$ within a triangular face in Fig.~\ref{fig:hyperkagome-unit-cell}, giving a unique PSG (given a choice of hyperkagome chirality\cite{Takagi2007}).

The mean field Hamiltonian $H_{\text{MF}}$ is a free Majorana bilinear Hamiltonian, so its ground state is immediately known by computing its band structure. The qualitative properties of this band structure carry the primary information, though the band structure energy scales contain the unknown mean field parameters $u$. The parameters $u$ can be determined self-consistently from the band structure by computing the Majorana propagator, as a Matsubara frequency integral of the inverse of the frequency-dependent Hamiltonian kernel. We have carried out the self consistency computation on the triangular and honeycomb lattice, using the Kitaev-type majorana flux-free PSG which is defined on these two lattices, and find that the mean fields evolve with $\alpha$ smoothly away from the Kitaev limit, with no first order transitions.

Regardless of the exact values of the mean field parameters, choosing the mean field to be analogous to the Kitaev honeycomb QSL already determines key properties of the resulting states on the various lattices.
First, all the lattices except for the honeycomb possess cycles with an odd number of bonds, such as triangles; this immediately requires the Kitaev Majorana mean field to spontaneously break time reversal symmetry\cite{Kivelson2007}. These time reversal broken spin liquids might not display typical characteristics of time-reversal broken states. For example, on the triangular lattice, even though time reversal as well as  $2\pi/6$ rotation each independently flip the flux pattern in triangular faces, the combined  operation of time reversal with $2\pi/6$ rotation
is still preserved as a single symmetry operation, so the Hall conductance vanishes. 
Second, lattices with an odd number of sites per unit cell necessarily have a spinon Fermi surface; the even-unit-cell lattices of pyrochlore and hyperkagome may or may not host gapped spinons. 

Third, certain qualitative features of the band structure are determined by the choice of mean fields, such as the consideration of only nearest neighbor bonds and the PSG. There are four Majorana fermion species per site; for a pure Kitaev Hamiltonian, $\chi^{1,2,3}$ have bands related to each other by the 120$^\circ$ SOC combined spin-spatial rotation, while $\chi^0$ has a generally different dispersion. For the honeycomb model, $\chi^0$ has a Majorana analogue of the Dirac cone, i.e. relativistic with zero mass, while $\chi^{1,2,3}$ all have completely flat bands separated from zero energy by a complete gap. The kagome lattice $\chi^{1,2,3}$ also has a flat band but it lies at zero energy i.e. at the Fermi energy, yielding the Fermi surface which necessarily arises here. The flatness results from a localized unpaired Majorana mode on one of the three sites in each unit cell; but since the remaining two sites form a line spanning the lattice, they disperse and the other bands are not flat, touching zero energy along lines in a quasi-1D spectrum. For the pyrochlore even qualitative statements cannot be currently made, since as mentioned above, there is no special choice of minimal flux PSG. On the hyperkagome with bond orientations as described above, $\chi^{0}$ has some gapless subextensively degenerate modes (such as from $\Gamma$ to $M$); but $\chi^{1,2,3}$, like for the honeycomb, have completely flat bands. These arise, as previously mentioned, because both the honeycomb and the hyperkagome fragment into extensively many disconnected clusters when only bonds of a single Kitaev label are kept. However, while the honeycomb clusters have an even number of sites and hence can form two fully gapped bands, separated from zero energy, the hyperkagome clusters have an odd (three) number of sites; each cluster always has one energy band at zero energy and hence $\chi^{1,2,3}$ are gapless.

\section{Outlook}
On the honeycomb lattice, the roles of the SU(2)-symmetric Heisenberg coupling and the SOC Kitaev coupling are distinct and clear: Heisenberg exchange yields magnetic order, Kitaev exchange yields the exactly solvable QSL phase. The natural interpolation between the two limits, that would occur if the couplings arise in iridium oxide compounds, is consistent with this framework: the intermediate region simply holds more magnetic order. However, as we have discussed above, generalizations of the Kitaev coupling naturally arise in iridium structures and other geometries of edge-sharing octahedra on many other lattices, motivating the study of the phase diagrams of \eqh on these various 2D and 3D lattices. Beyond the honeycomb, the roles of the two exchanges begin to break down.

The effect of lattice geometry on the ``frustration'' of a lattice is quite different for the two terms; the Hamiltonian and the lattice determine the frustration together, not independently.  More surprisingly, even in cases when the Heisenberg Hamiltonian appears highly frustrating, interpolating between the AF Heisenberg and the Kitaev limits, we find a phase which occurs on \textit{all} the lattices and which is exact and fluctuation free at a certain parameter point.
Subtle interplays of different magnetic couplings, rather than a monotonic ``frustration'' measure, seem to be at play.
The intermediate stripy phase is exact by virtue of being related to the ferromagnet, through a duality that emerges through the SOC on the $t_{2g}$ orbitals microscopically generating the Hamiltonian.

The Klein group structure of the mapping between Hamiltonians (a duality) is in some sense highly specific to these quantum chemistry considerations but in another sense, as a mathematical object $\mathbb{Z}_2 \times \mathbb{Z}_2$, quite universal. The duality transformation it generates is interesting here for another reason: while most dualities fix a single self-dual Hamiltonian and map the two regimes on either side of that point, with qualitatively different features, to each other, the Klein duality is different. It admits two self-dual Hamiltonians, which seem to generally lie in the interior of a phase. And it acts in a complicated way on spin and spatial indices, making its action as a $\mathbb{Z}_2$ symmetry operation highly nontrivial.

Regarding possible experimental significance of Hamiltonians arising from strong SOC, it is important to observe that the Kitaev couplings naturally occur in a manner more subtle and constraining than naive symmetry considerations would suggest:
 for example, Kitaev interactions can arise for iridium ions on the fcc but not on the simple cubic.
Computations of the quantum phase diagrams on the various lattices, especially the pyrochlore and hyperkagome, will pave the way towards  predictions and comparisons with experimental results.

\section*{Acknowledgements}
We thank Yi-Zhuang You, Yuan-Ming Lu, George Jackeli and Christopher Henley for useful discussions.
We are also grateful for the hospitality of the Kavli Institute for Theoretical Physics, where part of this work was written.
This research is supported in part by the NSF under Grants No. DGE 1106400 and NSF PHY11-25915 for the KITP Graduate Fellowship Program (IK), and ARO MURI grant W911NF-12-0461 (AV).

\begin{appendix}

\section{Details on the Luttinger-Tisza Approximation phase diagrams for the various lattices}
\label{sec:appendixLT}
Here we give the results of the Luttinger-Tisza Approximation (LTA) analyses in more detail.
The classical Kitaev-Heisenberg model on the triangular lattice has been recently studied\cite{Daghofer2012}, and the phase diagram for the honeycomb is already known\cite{Khaliullin2010}; we begin by reviewing the results for the triangular and honeycomb lattices for the sake of direct comparison with the other lattices.
Note that on all the lattices, the pure Kitaev points $\alpha=\pm 1$ always host degeneracies in the BZ; except where noted below, these are just subextensive degeneracies, which we expect will be lifted by small perturbations or by boundary conditions for a finite system.

\textbf{Triangular lattice:}
\textit{AFM Heisenberg, $\eta=+1$.} The pure Heisenberg antiferromagnet $\alpha=0$ hosts ordering at wavevector $K$ (BZ corner), corresponding to 120$^{\circ}$ order with a tripled unit cell. This ordering can also be seen by explicitly working with the enlarged unit cell. For $0 < \alpha < 1/5$ the ordering wavevector is incommensurate as it migrates from $K$ to $M$ (BZ edge midpoint). For $-1 < \alpha < 0$ the ordering wavevector is also incommensurate, moving from $K$ in the direction of $\Gamma$. Only one spin component has minimum energy for a given momentum here, preventing the construction of any order with unit length spins such as the usual coplanar spiral.
The antiferromagnetic Heisenberg limit 120$^{\circ}$ order actually has point topological defects (vortices), with  $Z_2$ character; the recent triangular lattice classical model study included classical Monte Carlo computations\cite{Daghofer2012} which found they may be stabilized into a vortex lattice in this incommensurate ordering phase.  Finally, for $1/5  < \alpha < 1$, we find stripy order at wavevector $M$.
\textit{FM Heisenberg, $\eta=-1$.} For $-1 < \alpha < 3/7$ there is ferromagnetic order. For $3/7 < \alpha < 1$ continues the phase of incommensurate ordering wavevector between $\Gamma$ and $K$ from $\eta=+1$.

\textbf{Honeycomb lattice:}
For pure Kitaev exchange $\alpha = \pm 1$ there are extensive degeneracies.
\textit{AFM Heisenberg, $\eta=+1$.} For $-1 < \alpha < 1/3$, simple Neel order (wavevector $\Gamma$) is stabilized, as the honeycomb is bipartite.  For $1/3 < \alpha < 1$ there is stripy order at $M$.
\textit{FM Heisenberg, $\eta=-1$.} Because we have a quadratic spin Hamiltonian and a bipartite lattice, this regime can be mapped to $\eta=+1$ by flipping all spin components on one of the two sublattices. For $-1 < \alpha < 1/3$ there is a ferromagnet,  for $1/3 < \alpha < 1$ there is zigzag order at $M$.

\textbf{Kagome lattice.}
\textit{AFM Heisenberg, $\eta=+1$.} The $\alpha=0$ Heisenberg antiferromagnet hosts extensive degeneracy. For $0 < \alpha < 1$ there is stripy order (wavevector $\Gamma$), here ferrimagnetic.
But while spins are uniformly normalized at the SU(2) point $\alpha=1/2$, away from $\alpha=1/2$ the energy is minimized when spins within the unit cell are of different lengths, suggesting the ordering becomes increasingly frustrated. For example near $\alpha=1/2$, if two of the spins within the unit cell are chosen to be normalized with magnitude $+1$, the third has magnitude $\frac{1-3 \alpha +\sqrt{9+\alpha  (-22+17 \alpha )}}{2 (-1+\alpha )}$, or roughly $-2+2\alpha$.
  For $-1 < \alpha < 0$ the LT method fails: two spins within a triangle plaquette align antiparallel along the axis of their bond label $\gamma$, but the third is so frustrated that its magnitude is set to exactly zero, also giving  spurious subextensive line degeneracies from $\Gamma$ to $M$.
\textit{FM Heisenberg, $\eta=-1$.}
$\alpha=0.5$, the Klein rotation of the Heisenberg antiferromagnet, again hosts extensive degeneracy. The ferromagnet exists over a wide range $-1 < \alpha < 1/2$ but with the same strong frustration as for the stripy phase: away from the Heisenberg limit $\alpha=0$, the energy is minimized by having spins of different lengths within the unit cell. For example, at $\alpha=-2/5$, the unit cell has all spins aligned but with one spin at $2/3$ the magnitude of the others.  The regime $0.5 < \alpha < 1$ continues the antiferromagnetic-Kitaev regime of $\eta=+1$, with two oppositely aligned spins and the third spuriously set to zero.

\textbf{Face centered cubic (fcc) lattice.}
\textit{AFM Heisenberg, $\eta=+1$.} $0 < \alpha < 1$ hosts the 3D-stripy order, antiferromagnetic along a cubic axis, which for the fcc BZ is at wavevector $X$ (center of one of three square faces of the BZ). For $X_z$ order, stripy-ordered spins align along $S^z$.s
For $-1 < \alpha < 0$, there is a line degeneracy from $X$ to $W$ (i.e. along diagonals of a square face of the BZ). The square face lies at some Euclidean direction, say $\hat{z}$. The two perpendicular spin components (here $S^x$ and $S^y$) participate at $X$, with one of the two stabilized as the wavevector moves towards a corner $W$. The set of minimal energy spin configurations, with normalized spins, therefore contains the wavevector $X$ stripy order though with spins uniformly rotated to the $S^x,S^y$ plane.
The orders at generic degenerate $q=(1-r)X+r W$ would usually form a spiral but here only have minimized energy for one of the three spin components, so they cannot stabilize any order with unit length spins.
For the constrained classical Hamiltonian and the quantum Hamiltonian, the higher energy spin component(s) are likely to be mixed into the ground state, so it appears the LT method hints at spiral order in this regime.
\textit{FM Heisenberg, $\eta=-1$.} The ferromagnet is stable for $-1 < \alpha < 1/2$. The region $1/2 < \alpha < 1$ is part of the spiral-like phase of $\eta=+1$, forming a single phase stabilized by  antiferromagnetic Kitaev interactions regardless of the sign or presence of Heisenberg exchange.

\textbf{Pyrochlore lattice.}
\textit{AFM Heisenberg, $\eta=+1$.} The $\alpha=0$ Heisenberg antiferromagnet hosts extensive degeneracy. Stripy order (wavevector $\Gamma$) is stable for $0 < \alpha < 1$. For $-1 < \alpha < 0$ the LT method fails: the preferred states have two of the four spins in the unit cell pointing antiparallel to each other, with the other two spuriously set to zero. The (subextensive) degeneracy extends across planes in the BZ, specifically the high symmetry planes containing the gamma point and the centers of (both  hexagonal and square) faces.
\textit{FM Heisenberg, $\eta=-1$.}  $\alpha=1/2$, the Klein rotation of the Heisenberg antiferromagnet, again hosts extensive degeneracy. The ferromagnet is stable for $-1 < \alpha < 1/2$. Strong antiferromagnetic Kitaev exchange at $1/2 < \alpha < 1$ continues the $\eta=+1, \alpha<0$ regime of plane degeneracies and two missing spins.  There are subextensive degeneracies at $\alpha=\pm 1$.
Note that unlike the kagome case, here even away from SU(2) points, the ferromagnet and stripy regimes have all spins of unit length.

\textbf{Hyperkagome lattice:}
\textit{AFM Heisenberg, $\eta=+1$.} The $\alpha=0$ Heisenberg antiferromagnet hosts extensive degeneracy, as do the two pure Kitaev limits $\alpha=\pm 1$. Stripy order (wavevector $\Gamma$) nominally exists for $0 < \alpha < 1$, but with the same strong frustrations (except for the SU(2) point $\alpha=1/2$) as for the kagome. For say $\hat{z}$ ordering, the four spins in the middle of their respective $z$-type clusters are chosen smaller(larger) than unity for $\alpha<1/2$($\alpha>1/2$).
The region with antiferromagnetic Kitaev exchange $-1<\alpha<0$ hosts a magnetically ordered phase which we term the \textit{cluster-ferrimagnet}.  In the $z$-type cluster-ferrimagnet, $z$-clusters all have the identical spin ordering ``(up, down, up)'', resulting in a state with a nonzero net magnetization.
The LTA gives the spins in the middle of each cluster a somewhat larger, unphysical, normalization, but this magnetic pattern is likely to be robust.
\textit{FM Heisenberg, $\eta=-1$.}  $\alpha=1/2$, the Klein rotation of the Heisenberg antiferromagnet, hosts extensive degeneracy like in the pyrochlore. (As already mentioned, the $\alpha=\pm 1$ points do as well.)
The ferromagnet over $-1 < \alpha < 1/2$ again has different length spins within the unit cell, except for the SU(2) Heisenberg point $\alpha=0$.  For $0.5 < \alpha < 1$ there is the \textit{cluster-antiferromagnet}, which is the Klein dual of the cluster-ferrimagnet. The two orders appear to be separated by the likely quantum phase, associated with extensive degeneracy, at $\eta \alpha=+1$. In this order, like for the cluster-ferrimagnet, the middle spin in each cluster points opposite to the other two and prefers to have a larger magnitude. Unlike the cluster-ferrimagnet, the cluster-antiferromagnet has no net magnetization since a pair of clusters is flipped relative to the other pair, as can be seen by the Klein duality.

\section{Mean fields and the projective symmetry group (PSG)}
The choice of partons (here Majorana fermions) with which to decompose the spin is not unique; there is actually full gauge SU(2) freedom in defining the parton operators. When parton mean field hopping and pairing terms gain nonzero expectation values, the freedom is partially lost so that the mean field Hamiltonian is only invariant under a global $\mathbb{Z}_{2}$ transformation. This is known as the invariant gauge group (IGG). Beyond the mean field level, the partons couple to a dynamical gauge field, with the gauge group of the IGG. Hence when IGG$=\mathbb{Z}_{2}$, the spin liquid is known as a $\mathbb{Z}_{2}$ spin liquid.

The mean field ansatz may be chosen to preserve as many of the symmetries of the original Hamiltonian as possible, say a set of symmetries $S$. While the mean field ansatz (the collection of mean field amplitudes on the lattice) may not be immediately invariant under a symmetry in $S$, we may choose to always attach an SU(2) gauge transformation to the symmetry, such that the mean field ansatz is invariant under their combination.
Relations of $S$ (operators that multiply to the identity) must have associated gauge transformations that multiply to an element of the IGG, ie a global $1$ or $-1$ sign in the case where IGG$=\mathbb{Z}_{2}$. This collection of attached real and gauge symmetry operations that leave the mean field Hamiltonian invariant is known as the projective symmetry group (PSG). It is a necessary part of the description of any mean field ansatz, and distinguishes different spin liquid phases even when they have the same gauge group.

\end{appendix}

\bibliography{IridatesCitations}

\end{document}